\makeatletter
\@namedef{ver@flushend.sty}{9999/12/31}
\makeatother
\pdfoutput=1

\documentclass[letterpaper,twocolumn,10pt]{article}
\usepackage{usenix2019_v3}

\usepackage{tikz}
\usepackage{amsmath}

\usepackage{tabularx}
\usepackage{tabulary}
\usepackage{array}
\usepackage{enumitem}
\usepackage{float}
\usepackage{enumitem}
\usepackage{booktabs}
\usepackage{makecell}
\usepackage{siunitx}

\usepackage[para,online,flushleft]{threeparttable}

\usepackage{color}

\hypersetup{%
  pdflang={en-US},
  pdfkeywords={privacy; user study},
  pdfdisplaydoctitle=true, %
  bookmarksnumbered,
  pdfborder = {0 0 0},
  colorlinks,
  linkcolor={black!80!black},
  citecolor={black!70!black},
  urlcolor={blue!70!black},
  pdfstartview={FitH},
  breaklinks=true,
  hypertexnames=false
}

\usepackage{titlesec}

\usepackage{versions}

\includeversion{arxiv}
\excludeversion{usenix}

\titlespacing{\subparagraph}{\parindent}{1.9pt}{1.5ex}
\titleformat*{\subparagraph}{\normalsize\em}

\begin{document}

\date{}

\begin{arxiv}
\title{\Large \bf Strategies and Perceived Risks of Sending Sensitive Documents\thanks{A version of this paper was presented at the 30th USENIX Security Symposium (Sec '21).}}
\end{arxiv}

\begin{usenix}
\title{\Large \bf Strategies and Perceived Risks of Sending Sensitive Documents}
\end{usenix}

 \author{
 {\rm Noel Warford$^{\ddagger}$, Collins W. Munyendo$^{ \mathsection}$, Ashna Mediratta$^{\ddagger}$, Adam J. Aviv$^{\mathsection}$, and Michelle L. Mazurek}$^{\ddagger}$\\
$^{\ddagger}$ University of Maryland, $^{\mathsection}$The George Washington University
 }

\maketitle
\definecolor{Orange}{rgb}{1,0.5,0}
\definecolor{Red}{rgb}{1,0,0}
\definecolor{Green}{rgb}{0,0.8,0.5}
\definecolor{Purple}{rgb}{0.75,0,1}
\definecolor{babypink}{rgb}{0.96, 0.76, 0.76}
\definecolor{BlueGreen}{rgb}{0.2,0.85,1}

\newcommand{\todo}[1]{\textsf{\textbf{\textcolor{Orange}{[[TODO: #1]]}}}}
\newcommand{\michelle}[1]{\textsf{\textbf{\textcolor{Green}{[[MM: #1]]}}}}
\newcommand{\adam}[1]{\textsf{\textbf{\textcolor{Purple}{[[AA: #1]]}}}}
\newcommand{\noel}[1]{\textsf{\textbf{\textcolor{BlueGreen}{[[NW: #1]]}}}}
\newcommand{\collins}[1]{\textsf{\textbf{\textcolor{babypink}{[[CM: #1]]}}}}
\newcommand{\revise}[1]{{\textcolor{black}{#1}}}

\newcommand{\new}[1]{\textcolor{Red}{#1}}

\newcommand{\surveyOne}[0]{Survey~1}
\newcommand{\surveyTwo}[0]{Survey~2}

\begin{abstract}
  People are frequently required to send documents, forms, or other materials containing sensitive data (e.g., personal information, medical records, financial data) to remote parties, sometimes without a formal procedure to do so securely.
  The specific transmission mechanisms end up relying on the knowledge and preferences of the parties involved. Through two online surveys ($n=60$ and $n=250$), we explore the various methods used to transmit sensitive documents, as well as the perceived risk and satisfaction with those methods. We find that users are more likely to recognize risk to data-at-rest after receipt (but not at the sender, namely, themselves). When not using an online portal provided by the recipient, participants primarily envision transmitting sensitive documents in person or via email, and have little experience using secure, privacy-preserving alternatives. Despite recognizing general risks, participants express high privacy satisfaction and convenience with actually experienced situations.
  These results suggest opportunities to design new solutions to promote securely sending sensitive materials, perhaps as new utilities within standard email workflows. 
\end{abstract}

 \section{Introduction}
\label{sec:intro}

Users are often required to send sensitive information --- such as personally identifiable information (PII), medical information, or financial information --- to remote parties. The approaches people use to send this information can vary based on personal skill level, available tools, the situational context in which this information is required, and, importantly, the perceived sensitivity of the data involved and the trust in the remote party receiving the data~\cite{ruoti2017weighing, wash2015too, woodruff2014would}.

Significant prior work has focused on why users do (or do not) adopt specific private communications channels, such as end-to-end encrypted messaging, as well as how to make these channels more usable and transparent~\cite{bai2016inconvenient,ruoti2016we,abu2017obstacles,de2016expert,ruoti2013confused}.
However, users who are required to send specific sensitive information to possibly unfamiliar recipients, perhaps in a new context, may not have the same tools at their disposal, or are  unaware of their availability or applicability.
Little is known about how or why people choose \emph{specific channels} for secure or private transmission of sensitive data.

In this paper, we explore how users cope when required to send sensitive information in the digital age. In particular, we sought to answer three key research questions:
\begin{enumerate}[noitemsep, label={\bf RQ\arabic*:}]
	\item What methods do people choose when sending sensitive information, and why?
	\item Are participants satisfied with their current approaches, particularly in terms of whether they offer sufficient privacy? Why or why not?
	\item What risks are people most concerned about when sending sensitive information?
\end{enumerate}

To address these questions, we conducted two online surveys. In the first survey (\surveyOne{}, $n=60$), we asked participants to provide primarily open-ended responses about the communication methods they used, or would expect to use, to send sensitive documents. We asked for responses to nine different scenarios, such as applying for a mortgage or an apartment, or opening a bank account. Participants reported on their satisfaction with the transmission methods, from both privacy and convenience perspectives, as well as their perception of potential risks and ways to mitigate these risks. Twenty participants responded to each scenario, and participants described 11 different methods, including delivering documents in person, physical mail, email, fax, and direct messaging.

We designed a second survey (\surveyTwo{}, $n=250$), containing predominantly closed-item questions with answer choices derived from \surveyOne{} responses. While \surveyOne{} was scenario-driven, \surveyTwo{} was method-driven. Participants identified at least one of eight most frequently cited methods from \surveyOne{} that they had used successfully to transmit sensitive information. They were then asked to describe a specific situation where one method was used successfully, followed by multiple-choice and Likert-type responses about the people and data involved and their satisfaction with the method. We also asked about privacy and risk, such as the comparative risk at the end-points or in transit.

We find that many participants typically deliver sensitive information using ``offline'' means --- most frequently in person, but also via physical mail and phone calls. Unsurprisingly, the most common digital approach is to use online forms or portals provided by the recipient; many participants also use standard (unencrypted) email. For a few \surveyOne{} tasks, such as sharing a password, there was a higher preference for direct messaging or phone calls, but little appetite for using secure technologies. \surveyTwo{} indicates that while participants have largely heard of secure technologies, relatively few have used them.
These results suggest that if a predetermined online form is not available or not appropriate, 
email is the only other widely used digital option. 
Nonetheless, in both surveys the vast majority of participants expressed satisfaction with both the convenience and privacy of their method.

Both surveys revealed interesting patterns in participants' perceptions of risks.
\surveyOne{} participants, answering open-ended questions, did not prioritize the risk to 
sensitive information \emph{in transit}, but rather \emph{what happens after it arrives}, 
either due to malicious action by the recipient or simply because the recipient did not take appropriate care with the data. When prompted with multiple choice questions, \surveyTwo{} 
participants weight in-transit risks and risks at the recipient similarly, but discount 
risk to the data at the sender (namely, themselves, e.g., whether their own email
storage is at risk).

Our findings illuminate opportunities to both improve end-user education and design new, transparent solutions for securely sending sensitive information. These tools could include building connections to secure-document-sending into existing communications modes like email, and improving retrospective privacy tools to help people delete sensitive content persisting at-rest once they are no longer needed.

 \section{Related Work}
\label{sec:related}

This paper builds on extensive research on secure communication. In 1999, Whitten and Tygar's classic paper, \emph{Why Johnny Can't Encrypt: A Usability Evaluation of PGP 5.0}~\cite{whitten1999johnny}, described numerous user-facing issues that make encrypted email impractical for many users, and  similar problems persist in  PGP 9.0~\cite{sheng2006johnny}. Follow-up research suggests that usability challenges in encrypted email continue~\cite{ruoti2016we, sweikata2009usability, garfinkel2005johnny, bai2016inconvenient, koh.bellovin.ea:easy}, despite many attempts to automate the process~\cite{ruoti2013confused}.

More recently, secure, end-to-end messaging applications (e.g., Signal, WhatsApp, Telegram) have proliferated as a straightforward and transparent way for users to communicate privately. Secure messaging adoption is driven largely by peer influence, rather than its security properties~\cite{de2016expert, abu2017obstacles}, and users may have misconceptions about the security properties, sometimes believing outside parties can read these encrypted communications, or that methods like SMS are \emph{more} secure~\cite{abu2018exploring, abu2017obstacles}. Inaccurate mental models of security may contribute to these misconceptions~\cite{wu2018tree}, and we also find that our participants do not strongly grasp secure communication.

A particular challenge in secure communication is how to indicate when
transmissions are (not) secure. Numerous researchers have investigated the effectiveness
of different indicators, including website authentication indicators~\cite{sunshine2009crying,bravo2013your,schechter2007emperor,felt2014experimenting} and phishing warnings~\cite{akhawe2013alice,petelka2019put}. Making these indicators intelligible and noticeable, without impeding workflows or engendering habituation, remains an open challenge.

Even after transmission, data may continue to reside on servers,  at the sender and/or receiver. Cloud storage poses a particular problem~\cite{snyder2013cloudsweeper} as many users have incorrect models of the longevity and location of cloud-based information. Clark et al.~\cite{clark2015saw} and Khan et al.~\cite{khan2018forgotten} found that when shown data stored in the cloud, most users find at least one item they wish to delete.
Users also lack urgency to delete cloud-stored information~\cite{ramokapane2017feel} and express interest in tools designed to do this~\cite{monson2018usability}.
This contradicts a common user perception that they have nothing to hide~\cite{solove2007ve}.
This problem is only amplified over time, with an increasing number of messaging platforms and other services relying on cloud storage and computing.

Difficulty protecting information may arise in part because users (and even experts) often have difficulty defining what information is sensitive in the first place~\cite{solove2007ve}.
Different users may also have different standards for what does or does not fall into the category of sensitive information, as this is highly reliant on context and personal preference~\cite{ruoti2017weighing,wash2015too,woodruff2014would}.

Researchers have also investigated how people learn about digital security and privacy, and how they develop associated behaviors. People's mental models for security often focus on direct and visible threats~\cite{wash2015too,Fulton:Entertainment}, and people tend to adopt behaviors based on where the behavior was learned, rather than its content~\cite{redmiles2016learned}. As might be expected, convenience-security tradeoffs also play an important role in adoption of security behaviors~\cite{Fagan2016Why}. Other work suggests that social factors, such as observing others performing a particular security behavior, can motivate users to take more security or privacy precautions~\cite{Das2014:SocialProof,Das2019:Triggers}. We observe that many of these factors --- convenience, social expectations, and directness of expected threat --- also play a role in our participants' choices when sending sensitive information.

 \section{Methods}
\label{sec:methods}

We designed and conducted two online surveys exploring participants' current
practices and perceptions related to sharing sensitive documents. In \surveyOne{},
$n=60$ participants commented on three (randomly selected from nine) scenarios
where they would need to communicate sensitive information or documents to
another party. Questions in this survey were primarily open-ended, in order
to obtain a wide range of responses about participants' experiences and
perceptions.

\surveyTwo{} builds on the results of \surveyOne{} with a
larger sample, $n=250$. Participants were randomly assigned
to answer questions about their experiences with and perceptions of one
transmission method they had successfully used.
\surveyTwo{} used primarily closed-item questions with answer choices drawn
from the qualitative analysis of \surveyOne{}.
Both surveys were approved by the University of Maryland IRB, and participants were recruited using Prolific.

\subsection{\surveyOne{}}
\label{sec:methods:surveyone}
\surveyOne{} consisted of four sections, described below.\begin{usenix}~\footnote{The full questionnaire is given in the extended paper (see Appendix~\ref{app:extend}).}\end{usenix}
\begin{arxiv} The full questionnaire is provided in Appendix~\ref{app:survey1}.\end{arxiv}
\begin{enumerate}[noitemsep]
\item {\em Instructions}: Participants were briefed about the purpose of the survey and provided consent. %
  \item {\em Scenarios (x3)}: Each participant was surveyed about three different, randomly chosen scenarios in which someone might send sensitive documents. For each scenario, participants were asked whether they had experienced the scenario before, and if so, how they provided the required information. Alternatively, participants were asked to {\em imagine} how they would transmit the information in such a scenario. We also asked, on a five-point Likert scale for each scenario,
  about their satisfaction with the communication method overall and in terms of privacy/security and convenience. %
  Each closed-item question had an open-ended followup question.

\item {\em Risks and Mitigation}: Participants were asked to identify and describe two risks (or concerns) with providing sensitive documents and two precautions they would take to reduce those risks, as well as if they have ever taken these precautions. All these questions were open-ended.
\item {\em Demographics}: Finally, we asked about demographics, including IT/CS background, income level, and experience working with a security clearance or in a sector with data privacy regulations (e.g., health care, law). Other demographic information was obtained directly from Prolific rather than via survey questions.
\end{enumerate}

\paragraph{Transmission scenarios}
We developed nine scenarios for sending sensitive documents based on vignettes used in prior work~\cite{woodruff2014would,abu2017obstacles} and  based on the authors' anecdotal experiences. Scenarios included: applying for a mortgage, sharing a password, participating in a background check (e.g., for volunteering with children), applying for an apartment, creating a checking account, sharing a password-protected document, enrolling a child in a new school, seeing a new doctor, and sending financial documents to a tax accountant.
Each participant viewed three scenarios, randomly selected
with counterbalancing, resulting in 20 participant responses per scenario.
Full text descriptions of each are provided in Appendix~\ref{app:scen}.

\paragraph{Updating for current events}
\surveyOne{} was administered in two rounds, before COVID-19 and after. In round two, as part
of the scenario section, we asked two additional questions about whether the participant had experienced the scenario before or after social distancing and whether social distancing had changed their (real or imagined) approach.

\subsection{\surveyTwo{}}
\label{sec:methods_survey_2}

We designed a second survey (\surveyTwo{})
to explore some of the results of \surveyOne{} in more detail. In contrast to \surveyOne{}, which was structured around scenarios,
participants in \surveyTwo{} were randomly assigned to answer questions about
particular transmission methods, later describing a {\em real} scenario in which they had used that method.\begin{usenix}~\footnote{The full questionnaire is given in the extended paper (see Appendix~\ref{app:extend}).}\end{usenix}
\begin{arxiv} The full questionnaire is provided in Appendix~\ref{app:survey2}.\end{arxiv}

We considered eight methods that participants
in \surveyOne{} commonly reported: email, direct messaging, in-person,
online form or portal, document sharing service (e.g., Dropbox), phone call, fax, and physical mail.
A participant would first identify which of these methods they had successfully
used to transmit sensitive information in the past. We also asked participants to specify any other unlisted methods they had used in a follow-up free-response question.

We then assigned the participant one of their ``successful'' methods, with the rest of the survey relating to that method.
Only methods that were included in our initial list were used for further questions, in order to ensure standardized and consistent questions across conditions.
We continually weighted the random assignment of successful methods toward less popular methods based on the results of \surveyOne{} and the current \surveyTwo{} recruitment in order to keep distribution among methods relatively even.

We asked participants to recall a specific scenario where they successfully used the
assigned transmission method and describe the
type of information sent, the recipient, and why this method was selected (e.g., did they or the recipient choose it?). These questions were closed-item, with answer choices based on common answers in \surveyOne{} and an option to write in an ``other'' response. As in
\surveyOne{}, we also asked about privacy and convenience satisfaction using a Likert scale.

We then asked other questions about privacy and risk, with answer choices also drawn
from themes we identified in \surveyOne{}. These included potential risks such as
a recipient revealing data by accident or on purpose, as well as interception in
transit. We also asked about whether the participant believed the recipient
could keep their data safe, whether the participant could themselves take action to keep
their data safe, and whether the information would be received by the intended
recipient. We asked about the likelihood of specific risks, including
reputational damage, physical harm, and identity theft. Finally, we asked
about the level of risk to the data at the sender, at the recipient, and in transit using Likert-type scales. Lastly, we collected the same demographics.

\begin{table}[t]
    \centering
    \small
    \caption{Reliability statistics for qualitative coding, including number of rounds required to reach agreement.}
    \label{tab:krippendorff}
    \smallskip
    \begin{tabular}{lrr}
        \toprule
        \textbf{Question} & \textbf{Rounds} & \textbf{Alpha} \\
        \midrule
        \textbf{\surveyOne{}} & & \\
        Methods used to send & 2 & 0.94 \\ %
        Satisfied with method & 2 & 1.00\\
        Satisfied with privacy & 1 & 0.93 \\ %
        Satisfied with convenience  & 1 & 0.86 \\ %
        Potential risks & 3 & 1.00 \\
        \midrule
        \textbf{\surveyTwo{}} & & \\
        What is being sent - Other & 1 & 0.82 \\
        Methods used to send - Other & 1 & 0.92 \\
        \bottomrule
    \end{tabular}

\end{table}

\subsection{Recruitment}

We recruited via Prolific, and participants
were required to reside within the U.S., have a 95\% approval on
Prolific, be at least 18 years old, and self-report fluency in
English. We used free-response questions to validate participants' answers
were on-topic and responsive, discarding only one potential participant
in \surveyOne{} and six in \surveyTwo{}.

We recruited $n=60$ participants for \surveyOne{} and $n=250$ for \surveyTwo{}. Participants were compensated \$4.00; \surveyOne{} took on average
17.4 minutes, while \surveyTwo{} averaged 11.7 minutes. \surveyOne{} data collection took place in early February and then May 2020, \surveyTwo{} data collection took place in November and December 2020.

\subsection{Data Analysis}
\label{sec:qualcoding}

For most open-ended answers (primarily but not exclusively \surveyOne{}),
we used an open-coding content analysis approach~\cite{krippendorff1989content}.
Two researchers worked together to develop an initial codebook for
each question, using 10\% of the provided answers. They then independently
applied the codebook to an additional 10\% of the data per round,
iteratively updating the codebook between rounds until strong reliability (Krippendorff's
Alpha $\geq0.8$) was obtained~\cite{krippendorff1970estimating, landis1977measurement}.
At that point, all data was recoded by a single coder using the final codebook.
Reliability values are given in Table~\ref{tab:krippendorff}. For open-ended
questions with 20 or fewer responses, this approach was impractical;
instead, two researchers coded each answer collaboratively.\begin{usenix}~\footnote{See extended paper (Appendix~\ref{app:extend}) for complete codebooks.}\end{usenix}
\begin{arxiv} Complete codebooks with  sub-codes are reported in Appendix~\ref{app:codebook}.\end{arxiv}

We pre-planned our quantitative analysis for \surveyTwo{}
around an ordinal logistic regression designed to
identify factors associated with privacy satisfaction
(on a five-point Likert scale)~\cite{mccullagh1980regression}.
We tested a range of potential covariates, selecting
a final model based on minimum Akaike Information
Criterion (AIC)~\cite{akaike1998information}. Complete details
are given in Section~\ref{subsec:rq2}.

For other comparisons of Likert-type variables, we
use Kruskal-Wallis H-tests to identify differences
among three or more items, followed by post-hoc Mann-Whitney U (MWU) pairwise-tests with the Holm-\v{S}id\'{a}k correction.

\subsection{Limitations}
\label{sec:methods_limit}
Our study has a number of limitations typical of exploratory survey research. First, data in \surveyOne{} was collected without the opportunity to ask follow-up questions (as would be the case in semi-structured interviews). As a result, some coded responses may not fully portray the nuances of participants' methods and perceptions.
To compensate, we designed \surveyTwo{} to validate those results with a larger population.

Free-response questions may suffer from satisficing, in which participants mention
the first item that comes to mind rather than answering comprehensively~\cite{Krosnick:satisficing}; participants who fail to mention something may simply not have included it, rather than explicitly disagreeing. As such, counts of participants should be considered a lower bound reflecting top-of-mind concerns, rather than absolute prevalence. \surveyOne{} also asked participants to imagine actions if a scenario was unfamiliar, which could also lead to satisficing, as well as other biases related to self-reporting and imagining hypotheticals. Our design for \surveyTwo{} sought to address this by only asking in depth about successfully used transmission methods, so that participants could report on real experiences instead of imagined ones.

Data collection in \surveyOne{} was bifurcated due to COVID-19, potentially biasing participant responses.
We added questions in the second round of \surveyOne{} addressing COVID-19 and found few differences, and so we did not focus
on COVID-19 effects in \surveyTwo{}.

There are inherent limitations in using crowdsourcing platforms like Prolific. Prior work has shown that these platforms provide reasonable samples for security- and privacy-relevant questions~\cite{redmiles2019well},
and Prolific has been shown to provide high-quality crowdsourced data~\cite{peer2017beyond}.

We focused only on U.S. participants, as we are most familiar with common data transmission
scenarios in the U.S. Our participant recruitment, as is generally the case for crowdsourcing platforms, tended to be more male and younger than the U.S. population as a whole. We neither expect nor claim the data to be fully representative; however, we believe we obtained a
reasonably broad view of transmission approaches and associated perceptions in the U.S. Future work could examine similar norms in other countries and cultures.

 \section{Results}
\label{sec:results_rq}

\begin{table}[t]
    \centering
    \small
    \caption{Demographics of participants in both surveys. Excludes ``no answer'' and ``prefer not to say'' options. ``Sensitive Information'' indicates whether a participant had encountered the listed types of information in a professional context.}
    \label{tab:demographics}
    \smallskip
    
    \begin{tabular}{llrrr}
    \toprule
     & & {\bf S1\#} & {\bf S2\#} & {\bf S2\%} \\
     \midrule
     {\bf Gender} & Female & 21 & 111 & 44.4\\
     & Male & 39 & 132 & 52.8\\
     & Non-binary & 0 & 7 & 2.8\\
     \midrule
     {\bf Age} & 18--30 & 38 & 84 & 33.6\\
     & 31--40 & 13 & 68 & 27.2\\
     & 41--50 & 6 & 35 & 14.0\\
     & 51--60 & 1 & 17 & 6.8\\
     & 61+ & 1 & 10 & 4.0\\
     \midrule
     {\bf Income} & < \$50K & 23 & 121 & 48.4\\
     & \$50K-\$100K & 24 & 86 & 34.4\\
     & > \$100K & 11 & 35 & 14.0\\
     \midrule
     {\bf Education} & No high school & N/A & 4 & 1.6\\
     & HS or equiv. & N/A & 72 & 28.8\\
     & Bachelor or associate & N/A & 121 & 48.4\\
     & Advanced degree & N/A & 52 & 20.8\\
     \midrule
     {\bf CS} & No & 45 & 190 & 76.0\\
     {\bf Experience} & Yes & 12 & 54 & 21.6\\
     \midrule
     {\bf Security} & No & 55 & 214 & 85.6\\
     {\bf Clearance} & Yes & 2 & 19 & 7.6\\
     \midrule
     {\bf Sensitive} & Credit card & 18 & 90 & 36.0\\
     {\bf Information} & HIPAA & 18 & 62 & 24.8\\
     & Social Security number & 17 & 85 & 34.0\\
     & FERPA & 6 & 20 & 8.0\\
     \bottomrule
     \end{tabular}
\end{table}

We first report on our participants, and then the results of both surveys. The results are  organized  by research question, as defined in Section~\ref{sec:intro}. %

\begin{table}[t]
    \centering
    \small
\caption{Distribution of participants across methods (\surveyTwo{}). Participants were randomly assigned one 
method among those they reported having used successfully.}
\label{tab:methods_distr_2}
\smallskip
    \begin{tabular}{lr}
    \toprule
    \textbf{Method} & \textbf{\#Part.} \\
        \midrule
        In person &  37 \\
        Online form/portal & 35 \\
        Email &  31 \\
        Physical mail & 32 \\
        Phone call & 34 \\
        Document sharing service & 27 \\
        Fax & 30 \\
        Direct messaging &  24 \\
        \bottomrule
    \end{tabular}
\end{table}

\paragraph{Participants}
\label{sec:results:participants}
Demographics for both surveys are provided in Table~\ref{tab:demographics} and are based on both self-reported data provided by Prolific and on direct questions from our survey. 
Most participants do not have a CS background (75\% \surveyOne{}, 76\% \surveyTwo{}) and few previously (or currently) have a security clearance (two and one participant in \surveyOne{} and \surveyTwo{}, respectively). 
Many describe having worked in roles where they may have handled sensitive information, e.g., Social Security numbers or health information. Participants skew younger and more male, as noted in Section~\ref{sec:methods_limit}.

Participants were evenly and randomly distributed among 
information-transmission scenarios in \surveyOne{}; 20 participants 
per scenario. 
We used 
frequency weighting to partially balance assignment to transmission methods in \surveyTwo{}. 
The distribution of participants across 
methods is given in Table~\ref{tab:methods_distr_2}.

\subsection{RQ1: What methods are used and why?}

\begin{table*}[t]
  \centering
  \caption{Transmission methods reported by participants in Survey 1, across scenarios.
  Counts are provided for all scenario instances, and broken down by real
  and imagined instances.
  Participants sometimes indicated more than one method per instance.} 
  \label{tab:methods}
  \smallskip
  \resizebox{\linewidth}{!}{
    \small
    \renewcommand{\arraystretch}{1.5}
\begin{tabular}{lrrrp{5.5cm}p{7.8cm}}
\toprule
 & \multicolumn{3}{c}{\textbf{Count}} \\
 \cmidrule{2-4}

	      \textbf{Code} & \textbf{Total} & \textbf{Real} & \textbf{Imag.} & \textbf{Description} & \textbf{Quote} \\ \toprule

        In person & 85 & 54 & 31 & Delivering the information by hand to the recipient, whether written down or simply told to them & ``I provided the information on an application in the office. It was on paper and when I was done I handed it to them'' \\

        Email & 52 & 24 & 28 & Sending the information via email, regardless of email platform or encryption & ``I sent the password to the persons private email that I knew for a fact was only accessible by only them.'' \\

        Online form or portal & 32 & 30 & 2 & Using an institution's site, app, or portal to upload the information & ``I applied for a savings account recently, but I did it through their mobile app. Really they had all of my information, but they did ask to confirm questions like social security number and contact information.'' \\

        Direct messaging & 22 & 14 & 8 & SMS, secure and insecure messaging services, and other similar modes of communication & ``I would text them the password and tell them to delete it from their phone after they are done.'' \\

        Phone call & 19 & 9 & 10 & A direct telephone call & ``I provided it to him over a phone call.  I do not trust electronic devices with password sending.'' \\

        Fax & 8 & 3 & 5 & Faxing documents to the recipient & ``I would fax the documents to them simply because I do not trust sending that information via the internet.'' \\

        Sending online (unspec.) & 8 & 6 & 2 & Sending the information online without providing a specific method beyond that. & ``I would send all documents online.'' \\

        Physical mail & 8 & 7 & 1 & US Postal Service, UPS, Fedex, and other services & ``I would probably print everything out and snail mail it all to the doctor.'' \\

        Secure sending online & 6 & 5 & 1 & As ``Sending online'' above, but with an indication of security while simultaneously remaining nonspecific & ``Sending it securely online is a more convenient way to do it for everyone involved.'' \\

        Document sharing service & 3 & 2 & 1 & Services like Google Drive, Box, or Dropbox where a document is uploaded to a shared location & ``Maybe through an app like dropbox with both password and PDF in a shareable link.'' \\

        Video call & 1 & 1 & 0 & Facetime, Google Meet and other similar platforms & ``I think the best way would be through what i already described being email or webcam call or text or a secure form to submit to them.'' \\ \bottomrule

      \end{tabular}}

\end{table*}

\begin{table}[t]
  \caption{Participants' reasons for being ``satisfied'' or ``very satisfied'' with privacy of their transmission modes in Survey 1.
  Participant answers may have had more than one code.}
  \vspace{+.2in}
  \centering
  \small
\begin{tabular}{lr}

 \toprule
{\bf Satisfied Privacy Response Code} & {\bf Frequency} \\
  \midrule
    My method of sending is secure &  62 \\
    The recipient will keep my information safe &  34 \\
 	Information received by the intended recipient &  21 \\
    I am satisfied (no specification) &  12 \\
    I am unsure about the security of my method &   7 \\
    The method of sending is insecure &   6 \\
    I can keep my information safe &   5 \\
    I am unsatisfied &   2 \\
    The recipient may unintentionally disclose  &   2 \\
   \bottomrule
\end{tabular}%
\label{tab:whysatisfied_priv}
\end{table}

\begin{figure}[!t]
\includegraphics[width=\columnwidth]{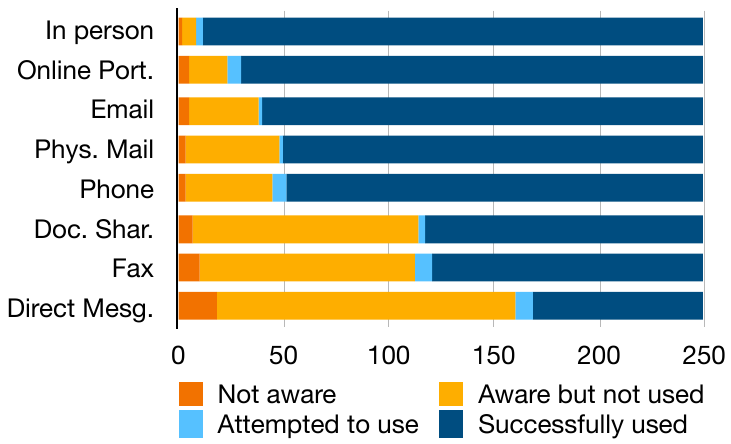}
\caption{Methods previously used by participants to send sensitive information (\surveyTwo{})}
\label{fig:which_methods_2}
\end{figure}

\paragraph{\surveyOne{}}
In \surveyOne{}, participants provided open-ended answers 
about how they sent required information in different scenarios. 
If they had not experienced the scenario, we asked them to 
imagine how they would send information in the scenario. We refer to these as \emph{real} and \emph{imagined} responses, respectively. Table~\ref{tab:methods} describes each identified mode of transmission,
as defined in our codebook, with frequency of occurrence across all scenarios.
(Table~\ref{tab:scene} in Appendix~\ref{app:figstables} provides the most common transmission methods per scenario for both real and imagined instances.)

By far, the most commonly reported transmission methods were taking the documents in person and sending the documents via email, especially for imagined scenarios.
Online forms, direct messages, and phone calls were also common methods, and
some responses indicated non-digital transmission methods. For example, P25 made an unprompted reference to not trusting digital methods: ``I would fax the documents to them simply because I do not trust sending that information via the internet.''

From \surveyOne{} responses, we wondered whether these methods, which were clearly top of mind, were also the methods participants had the most experience with. We also wondered whether participants knew about certain modes but had chosen not to use them, or were unfamiliar with them at all. Responses also 
suggested that participants frequently used methods chosen (or required) by the recipient, rather than choosing the method 
independently. We addressed these questions as part of \surveyTwo{}.

\paragraph{\surveyTwo{}}

In \surveyTwo{}, participants were asked whether they had 
used or heard of the most commonly described eight transmission methods from 
\surveyOne{}. The results, shown in Figure~\ref{fig:which_methods_2}, indicate that most 
participants were aware of most methods, clarifying an uncertainty in \surveyOne{}. Participants were most successful using in person, email, and online forms, aligning with the findings of \surveyOne{}, and
also mirroring \surveyOne{}, document-sharing services 
(e.g., Dropbox or Google Drive) and faxing were relatively uncommon. 

There are differences between \surveyOne{} and \surveyTwo{}.
Physical mail was rarely mentioned in 
\surveyOne{}, but participants had high levels 
of experience with it in \surveyTwo{}; on the other hand, while direct 
messaging was relatively popular in \surveyOne{}, it was 
the least frequently used method in \surveyTwo{}. (Participants also provided other methods used; the resulting qualitative codes appear in Table~\ref{tab:other_methods_2}, in the Appendix~\ref{app:figstables}.)

\begin{figure*}[t]
\centering
\includegraphics[width=0.98\textwidth]{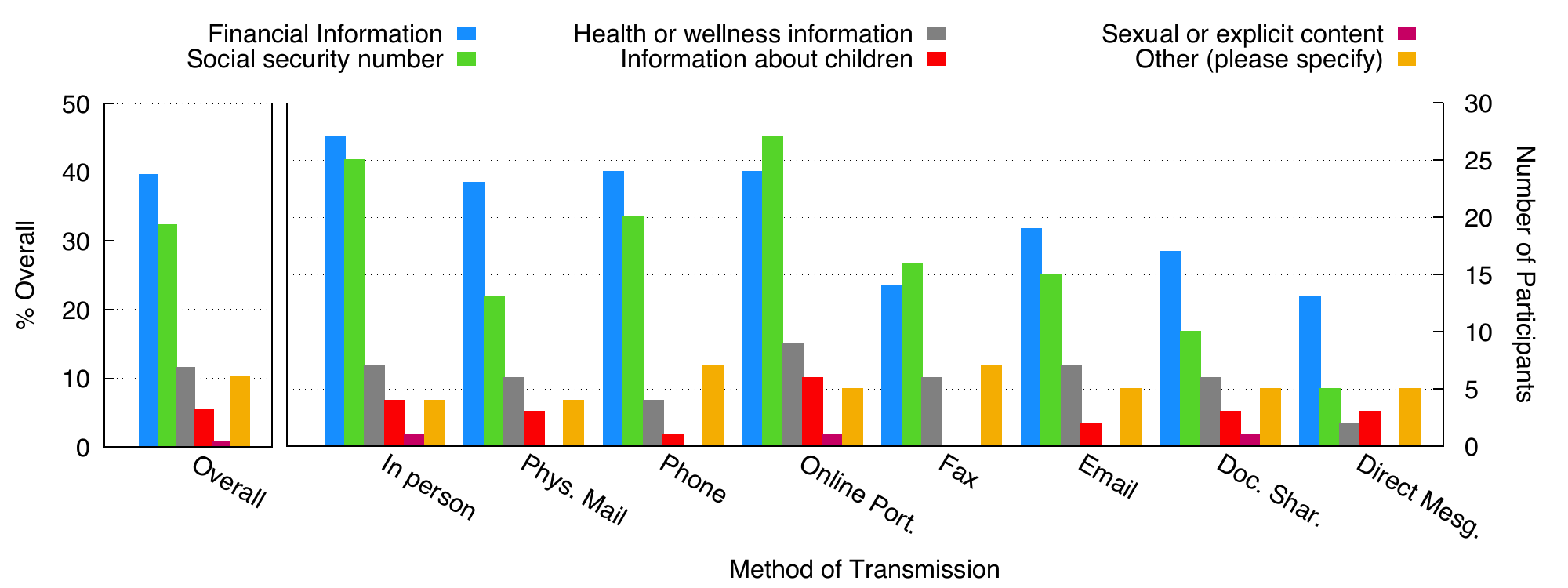}
\vspace{-.2in}
\caption{Type of information sent by participants across different methods (\surveyTwo{})}
\label{fig:what_sent_method_2}
\end{figure*}

\surveyTwo{} participants were randomly assigned a successfully used  transmission 
method, with counter-weighting for balancing (see Table~\ref{tab:methods_distr_2}). Participants were asked to recall an instance of using their assigned 
method to send sensitive information 
and reported sending many types of 
information (Figure~\ref{fig:what_sent_method_2}), with 
financial information and Social Security numbers (SSNs) most
common top-of-mind instances.
``Other'' responses, aggregated via open coding, are 
also weighted heavily toward identifying information. 
(These are summarized in Table~\ref{tab:what_sending_other_2} in Appendix~\ref{app:figstables}.) There is little variation in what was being sent for each transmission method (Figure~\ref{fig:what_sent_method_2}).

\begin{figure*}[t]
\centering
\includegraphics[width=0.98\textwidth]{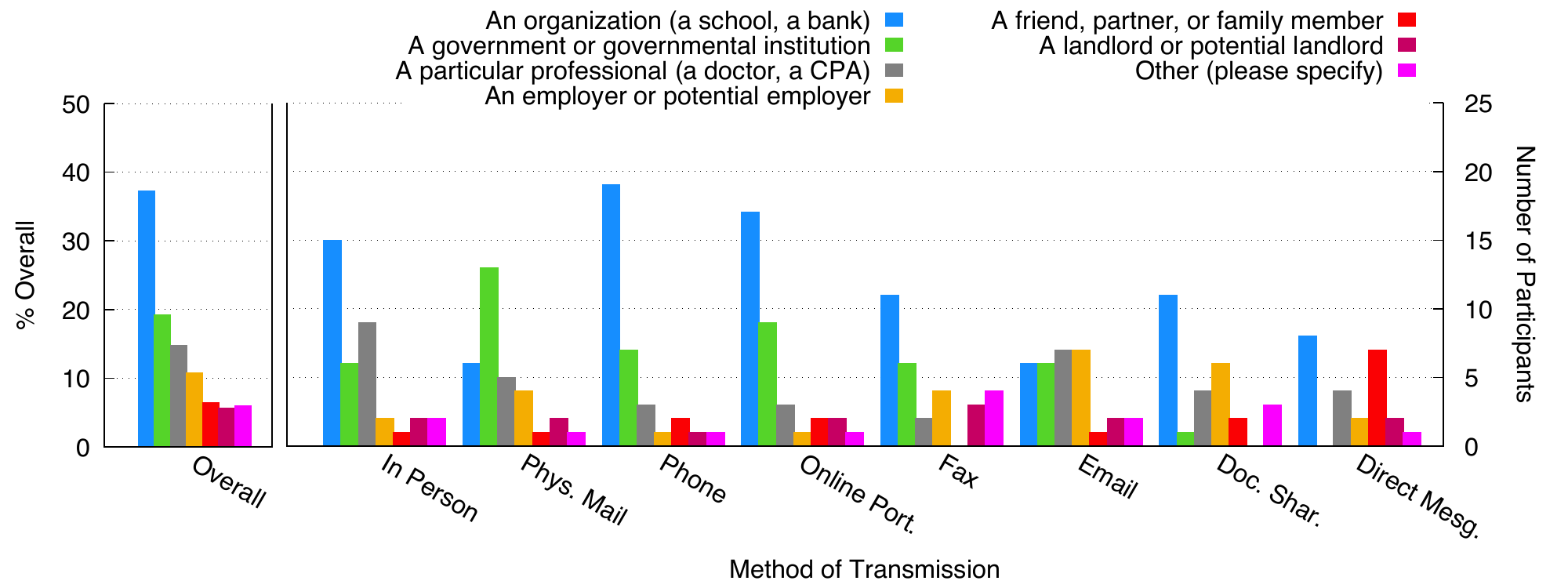}
\vspace{-.2in}
\caption{Recipients of sensitive information across different methods (\surveyTwo{})}
\label{fig:what_recipient_method_2}
\end{figure*}

We also asked about the recipient of the information, 
in categories including an organization (e.g., a bank), 
a particular professional (e.g., an accountant), 
a friend or family member, and others. Results are shown 
in Figure~\ref{fig:what_recipient_method_2}. In keeping with the trend 
toward financial and identity information, the most 
popular responses were an organization, a government 
agency or institution, and individual professionals. 
Governments received mail most often, and direct messages often went to friends and family.
Governments and organizations, unsurprisingly, were also most likely to use an online form.


\begin{figure*}[t]
\centering
\includegraphics[width=0.9\linewidth]{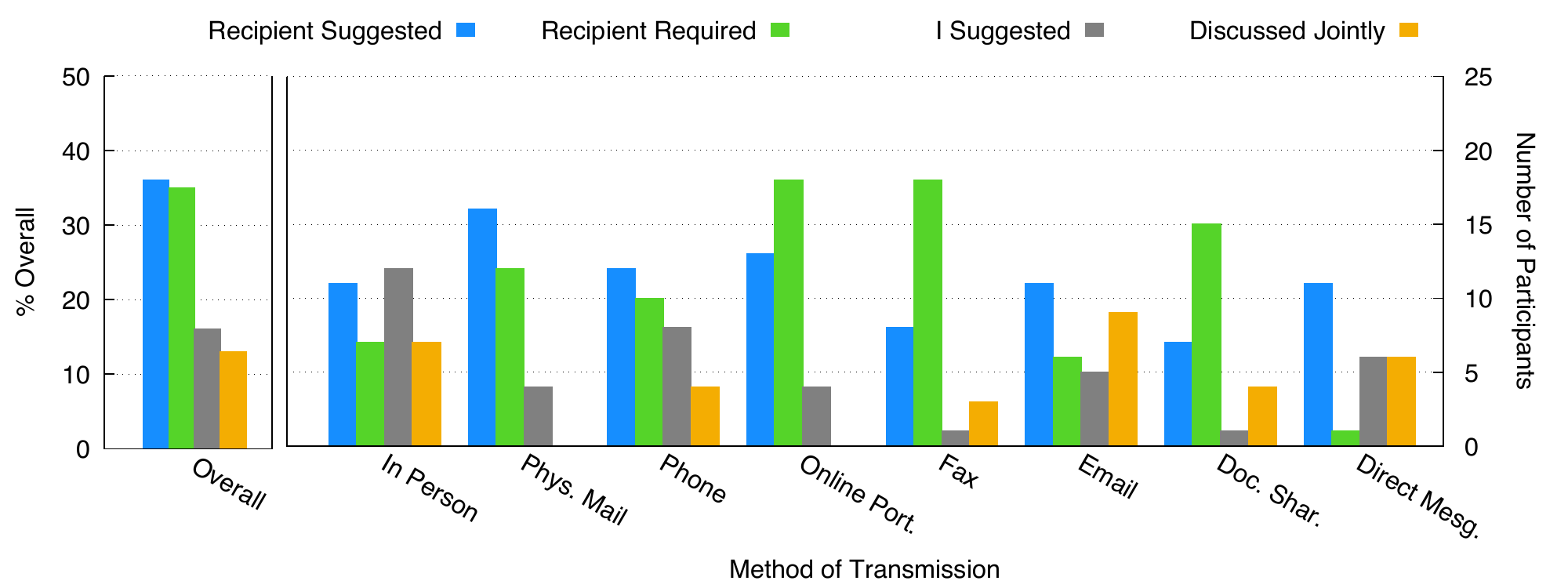}
\vspace{-.2in}
\caption{Determinants of the methods used to send sensitive information (\surveyTwo{})}

\label{fig:who_chose_t_method_2}
\end{figure*}

We also asked participants if they or the recipient 
chose the method (Figure~\ref{fig:who_chose_t_method_2}, left). 
Overwhelmingly, participants indicated that the recipient 
had either suggested or required the method.
Recipients required or suggested faxes, online forms, physical mail, and document sharing services, while participants more frequently suggested taking the documents in person or via phone call.
When discussing jointly, participants and recipients often landed on email (Figure~\ref{fig:who_chose_t_method_2}, right). 

\paragraph{Key findings for RQ1}
Email, online forms and taking documents in person are the most common transmission methods. Fax, document sharing services like Google Drive or Dropbox, and direct messages are least common. Participants have heard of these less common methods but not used them as frequently. 
Recipients are more likely than senders to choose the method of transmission.

\subsection{RQ2: Are people satisfied? Why?}
\label{subsec:rq2}

\begin{figure*}[ht]
    \includegraphics[width=\textwidth]{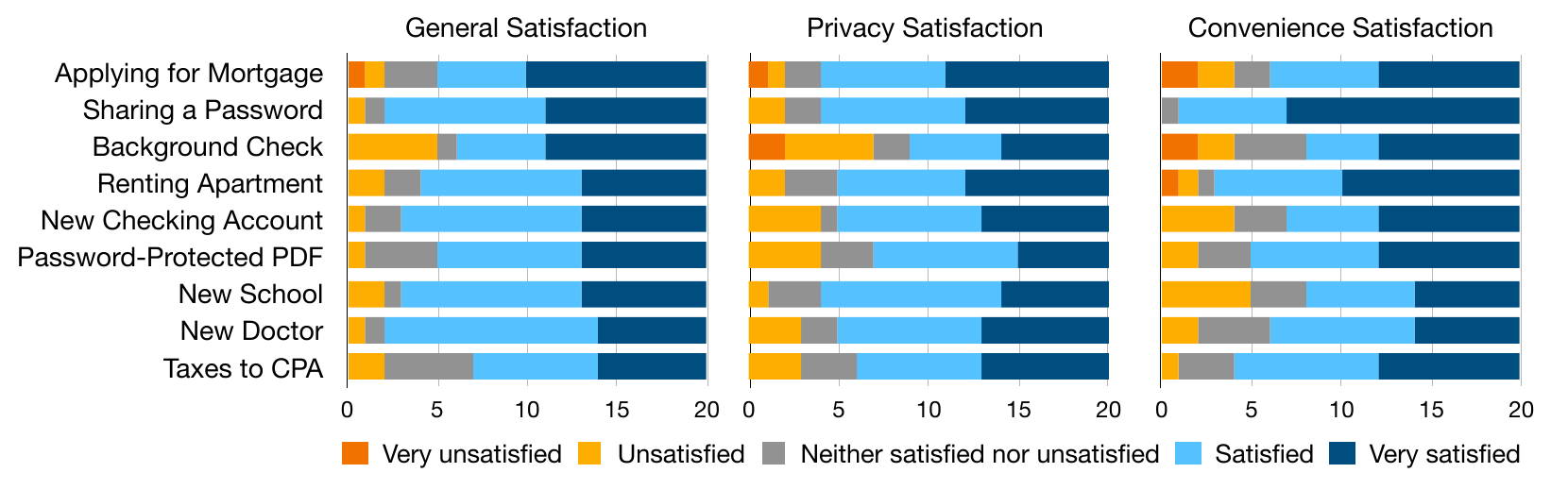}
    \vspace{-.2in}
    \caption{Satisfaction levels across different scenarios (\surveyOne{})}
    \label{fig:scen_sat_1}
\end{figure*}

\paragraph{\surveyOne{}} Most participants in 
\surveyOne{} were generally satisfied with the privacy and convenience of the transmission methods, and this satisfaction was consistent across scenarios (see Figure~\ref{fig:scen_sat_1}).
We asked participants to describe why they were satisfied (or unsatisfied) with the method's privacy (see Table~\ref{tab:whysatisfied_priv}), and many (62 instances) described satisfaction due the security of the method.
Another common reason for satisfaction (34 instances) laid at the communication endpoint. Participants believe the receiver will maintain security  and privacy, and thus they are satisfied with the privacy of the transmission method.

Participants who reported being ``unsatisfied'' or ``very unsatisfied'' with the privacy of their transmission method expressed concern that the method being used is insecure (12 occurrences), or mentioned dissatisfaction in general without specifying further (6 occurrences).
Some participants clearly described their distrust in a method but not why or how a threat might arise. For example, one participant mentioned the threat of ``access by others,'' but not how or why this would happen.
To explore this topic further, the codes from these free responses were used to develop Likert-scale questions about satisfaction.\begin{usenix}~\footnote{See \surveyTwo{} questions 14--15 in the extended paper (see Appendix~\ref{app:extend}).}\end{usenix} \begin{arxiv} (See questions 14 and 15 in Appendix~\ref{app:survey2}.)\end{arxiv}

\paragraph{\surveyTwo{}}

\begin{figure*}[t]
\includegraphics[width=\linewidth]{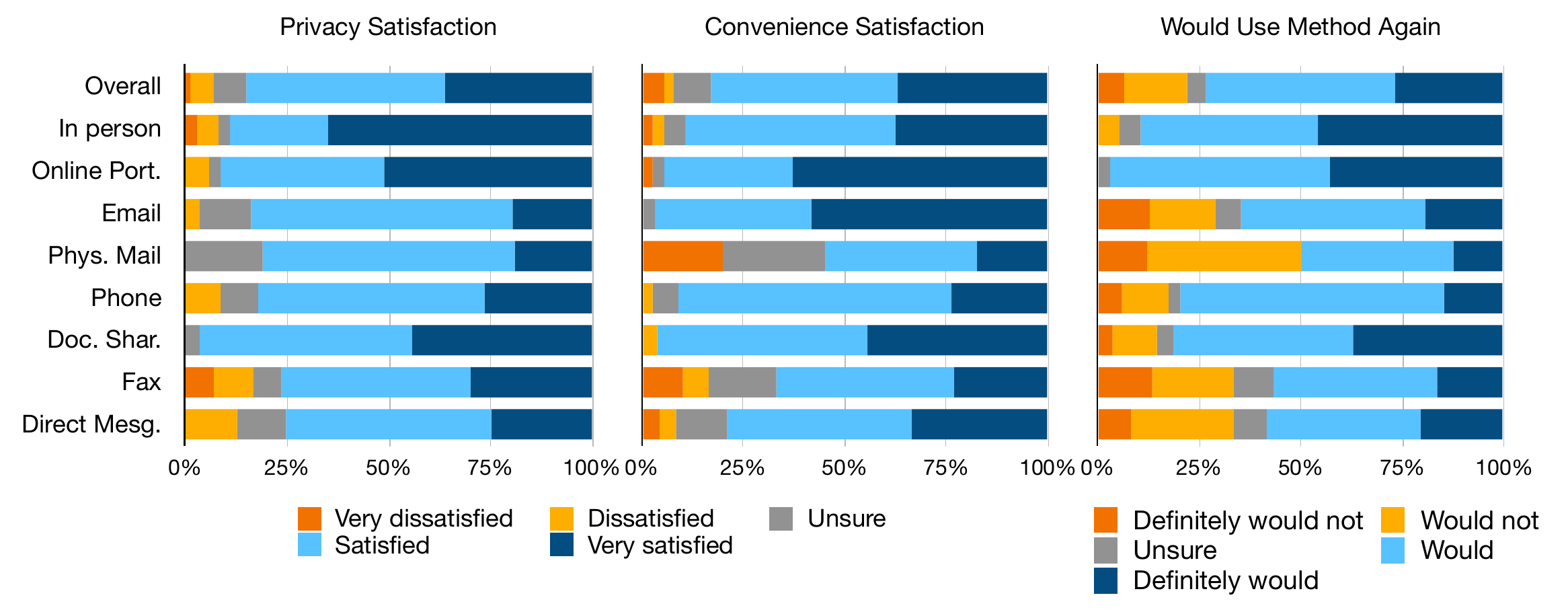}
\vspace{-.1in}
\caption{Satisfaction levels and willingness to use different methods again (\surveyTwo{})}
\label{fig:priv_conv_sat_would_use_overall}
\end{figure*}

\begin{figure}[ht]
\includegraphics[width=\columnwidth]{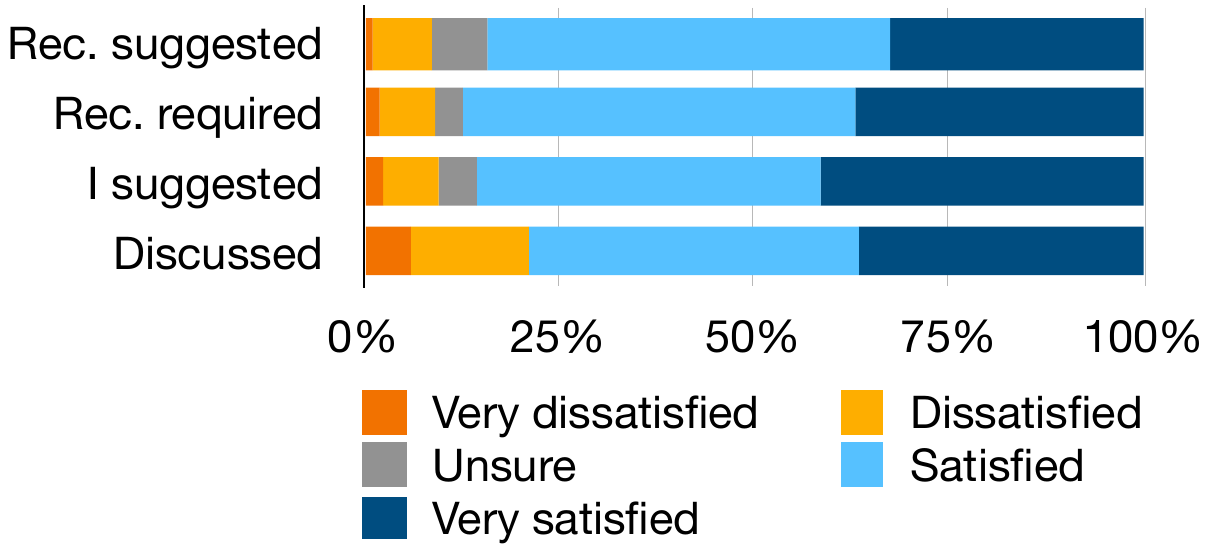}
\vspace{-.2in}
\caption{Privacy satisfaction based on the determinant of the method used (\surveyTwo{})}
\label{fig:privsat_choice_2}
\end{figure}

\begin{figure}[]
\includegraphics[width=\columnwidth]{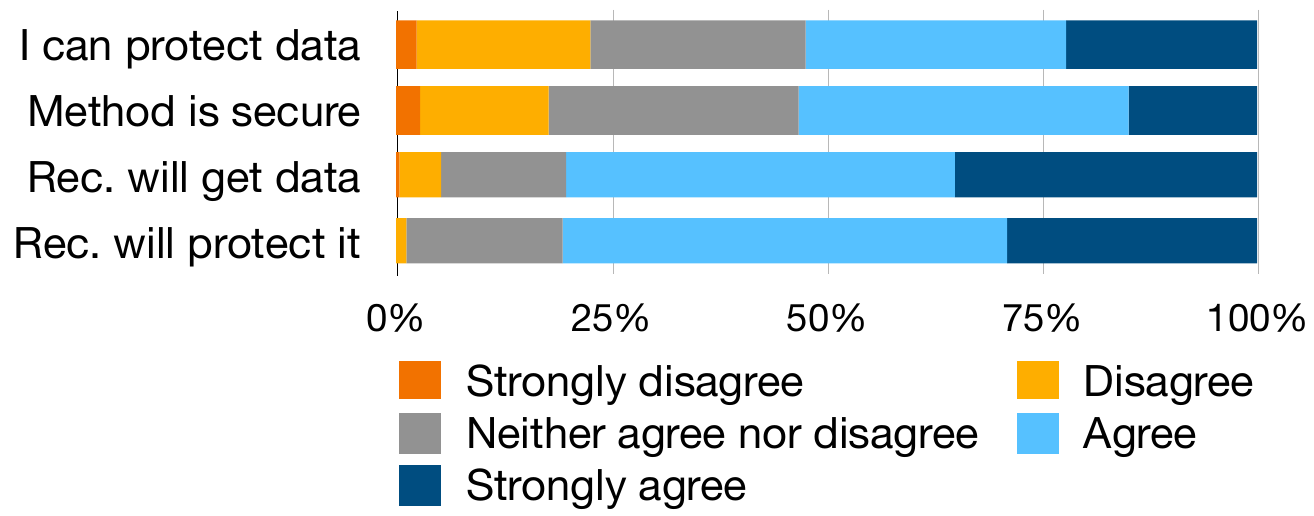}
\caption{Agreement with reasons for privacy satisfaction across all methods (\surveyTwo{})}
\label{fig:agreement_2}
\end{figure}

In \surveyTwo{}, we again observed that a large majority of participants were satisfied with the privacy of their methods (Figure~\ref{fig:priv_conv_sat_would_use_overall}, left). Only online forms, taking the documents in person, fax, and email registered any (and very few) ``Very dissatisfied'' responses. 

\begin{table}[t]
 \centering
 \footnotesize
  \caption{Final selected ordinal logistic regression model for participants' privacy dissatisfaction. Odds ratios above 1 indicate more dissatisfaction, relative to the baseline. The baseline for method is  ``taking the documents in person''; 
  other baselines are false, disagree, and unlikely. Pseudo-$R^{2}$: 0.55}
  \label{tab:regression2}

  \begin{tabular}{llrcr}
\toprule
                & & \textbf{Odds}  & \textbf{Conf.}  & \\
\textbf{Variable} & \textbf{Value}  & \textbf{Ratio} & \textbf{Int.}   & \textbf{\textit{p}-value}\\
\midrule
{\bf Method} & \emph{In person} & --- & --- & --- \quad \quad \\
 & Online form & 1.2 & [0.4, 3.4] & 0.761\phantom{*} \\
 & Email & 3.7 & [1.3, 11.3] & 0.017* \\
 & Mail & 3.8 & [1.4, 10.9] & 0.012* \\
 & Phone & 3.0 & [1.1, 8.5] & 0.034* \\
 & Doc sharing & 1.1 & [0.4, 3.4] & 0.876\phantom{*} \\
 & Fax & 3.4 & [1.2, 9.9] & 0.026* \\
 & DM & 1.9 & [0.6, 6.3] & 0.274\phantom{*} \\
\midrule
{\bf Financial} & True & 1.9 & [1.1, 3.6] & 0.032* \\
{\bf SSN} & True & 0.5 & [0.3, 0.8] & 0.007* \\
{\bf Health} & True & 1.1 & [0.5, 2.2] & 0.891\phantom{*} \\
{\bf Children} & True & 1.0 & [0.4, 2.6] & 0.978\phantom{*} \\
{\bf Explicit} & True & 4.4 & [0.4, 39.0] & 0.178\phantom{*} \\
\midrule
{\bf Risk at dest.} &  Agree & 1.9 & [1.1, 3.3] & 0.032* \\
{\bf Recip't keep safe} & Agree & 0.4 & [0.2, 0.7] & 0.006* \\
{\bf Method secure} & Agree & 0.2 & [0.1, 0.3] & < 0.001* \\
{\bf Recip't share }\\
{\bf \phantom{xx}on purpose} & Likely & 2.8 & [1.0, 8.4] & 0.056\phantom{*} \\
\bottomrule
\end{tabular}
\end{table}

\subparagraph{Regression on privacy satisfaction}
We ran an ordinal logistic regression (our main planned analysis) to see what factors most correlated with privacy satisfaction when sending sensitive information. We report the results in Table~\ref{tab:regression2}.
Privacy satisfaction was our outcome variable.
Potential covariates included the following:
\begin{itemize}[noitemsep]
    \item Method used
    \item Type of data
    \item Identity of the recipient
    \item Level of trust in the recipient
    \item Who chose the transmission method
    \item The reported tech-savviness of both the participant and the recipient
    \item Likert-type responses for a variety of items, generated based 
    on \surveyOne{} free responses. Responses were binned into binary variables for analysis.\begin{usenix}\footnote{These statements are slightly abbreviated; full text can be found in questions 14, 15, and 21 in the extended paper (see Appendix~\ref{app:extend}).}\end{usenix}
    \begin{arxiv}\footnote{Full text for these items can be found in questions 14, 15, and 21 as reported in Appendix~\ref{app:survey2}. The statements here are slightly abbreviated.}\end{arxiv}:
    \begin{itemize}[noitemsep]
        \item The recipient will unintentionally reveal my data. 
        \item The recipient will intentionally reveal my data.
        \item My data will be intercepted in transit. 
        \item The recipient can keep my data safe.
        \item I can do something to keep my data safe.
        \item This method is inherently secure. 
        \item The information would be received as intended.
        \item The data is at risk on my end.
        \item The data is at risk in transit.
        \item The data is at risk at its destination.
    \end{itemize}
\end{itemize}

Definitions and levels for the above factors can be found in Table~\ref{tab:factor_levels} in Appendix~\ref{app:figstables}.

We used a Variance Inflation Test (VIF) to check multicollinearity in the initial model with all of the above factors. All variables were well below the threshold value of 5 except for the ``Other (please specify)'' option for the type of data being sent. Since the types of data being sent were each independent binary factors, rather than a single categorical choice, we excluded this factor from our model selection process.

We then compared a set of potential models, keeping method used and 
type of data (except for the factor we removed) in every model but testing all possible combinations of the other covariates, not including interaction factors. We excluded interaction factors because we did not have sufficient power to include all the potential combinations.
For parsimony, we selected the model with 
minimum Akaike Information Criterion (AIC)~\cite{akaike1998information}. 
The final model, shown in Table~\ref{tab:regression2}, exhibits 
a pseudo-$R^{2}$ of 0.55 using the Aldrich-Nelson method, as evaluated by Hagle and Mitchell~\cite{hagle1992goodness}, indicating a 
fairly strong fit.
Odds ratios above 1 indicate an increase in dissatisfaction 
relative to the baseline, as dissatisfaction was much less 
common than satisfaction.
The model identifies several covariates 
that significantly correlate with privacy dissatisfaction, as follows.

\subparagraph{Transmission methods}
Relative to the baseline of in-person transmission --- selected because 
it is the only method that does not require communications infrastructure 
--- physical mail is associated with a 3.8$\times$ higher likelihood 
of more privacy dissatisfaction.\footnote{We note that this sample was 
collected in the U.S. shortly after the 2020 presidential election, 
during which the reliability and security of the postal service 
received significant negative attention.}
Email, phone calls, and faxes similarly exhibited odds ratios 
greater than or equal to 3. No other method was significantly different from 
in-person. 

\subparagraph{Type of data}
Several types of data being transmitted were also significantly 
correlated with privacy dissatisfaction. Because participants 
were allowed to select multiple potential options for data type, 
data types are modeled in the regression as independent boolean 
factors (baseline is false). Participants reported significantly 
more privacy dissatisfaction (odds ratio: 1.9) when financial 
information was included in the transmission. Surprisingly, they reported less dissatisfaction (odds ratio: 0.5) when transmitting 
Social Security numbers. This effect appears to be driven by 
an unusually large number of participants reporting ``very satisfied'' 
for transactions involving Social Security numbers.

\subparagraph{Likert factors}
Figure~\ref{fig:agreement_2} illustrates responses to some 
of the Likert-type questions relating to reasons for 
privacy satisfaction (questions based on \surveyOne{} 
responses). On the whole, participants were confident 
recipients would receive and protect data but 
less confident that transmission methods were secure or 
that they themselves could protect data. 

Four Likert-type statements appear in the final regression model 
for privacy satisfaction: 
agreeing/disagreeing that the data is at risk at the 
destination, that the recipient can keep data safe, and 
that the method is inherently secure; as well as likelihood 
that the recipient will intentionally reveal data. 

Participants were $1.9\times$ as likely to report more privacy 
dissatisfaction when they agreed that data was at risk at the 
destination. In contrast, participants reported lower 
dissatisfaction when they agreed the recipient could keep their data 
safe (odds ratio: 0.4) or agreed the method was inherently secure 
(odds ratio: 0.2). All of these results are intuitive and match 
participants' comments from \surveyOne{}.

\subparagraph{Other factors}
None of the other factors we tested appeared in the final model, 
indicating that they are not meaningfully correlated with privacy 
satisfaction. Somewhat to our surprise, these non-factors included 
whether the recipient or the participant chose the method; this 
result is illustrated in Figure~\ref{fig:privsat_choice_2}.

\subparagraph{Convenience and Reuse}
We also asked participants whether they were satisfied overall 
with the convenience of their method, and whether they would 
use the method again.

Much like \surveyOne{}, large majorities of participants 
were satisfied with convenience (see Figure~\ref{fig:priv_conv_sat_would_use_overall}, 
center). Post-hoc, pairwise MWU comparisons (see  Section~\ref{sec:methods_survey_2}) indicate 
participants found physical mail significantly less convenient 
than in-person, online portal, email, and document sharing, 
and found faxing significantly less convenient than online portals or email.
(Full details are given in Table~\ref{tab:appendix:kw-convenience} in Appendix~\ref{app:figstables}.)

Despite the overall satisfaction with privacy and convenience, we saw 
somewhat more variance when the participants were asked if they would 
use the method again (Figure~\ref{fig:priv_conv_sat_would_use_overall}, right). 
Post-hoc, pairwise MWU comparisons find that participants were most 
likely to want to use an online portal again (significantly more than 
email, physical mail, phone, fax, or direct messages). In-person was 
also significantly more popular for reuse than physical mail or fax. 
(Full details are given in Table~\ref{tab:appendix:kw-woulduse} in Appendix~\ref{app:figstables}.)

\paragraph{Key findings for RQ2}
Participants are overwhelmingly satisfied with the privacy of their methods, even when they did not choose the transmission method. Reasons for this largely depend on the \emph{recipient} keeping data safe 
as well as confidence in the inherent security of their method. 
Both taking the documents in person and using an online 
portal --- despite seemingly being quite different from each other --- 
are perceived as providing a good overall tradeoff among privacy and convenience.

\subsection{RQ3: What risks are people most concerned about?}
\label{subsec:rq3}

\paragraph{\surveyOne{}}

\begin{table}[t]
    \centering
    \small
\caption{Perceived risks of sending sensitive documents in Survey 1.
Participant answers may have contained more than one code.}
\label{tab:risks_abbr}
	\smallskip
   \begin{tabulary}{\columnwidth}{Lr}
        \toprule
        \textbf{Risk} & \textbf{Frequency} \\ \midrule
        The data at rest is at risk & 50 \\
        Unspecified ``malicious intent'' & 25 \\
        Identity theft & 15 \\
        The data in transit is at risk & 12 \\
        The data will be lost or misplaced & 9 \\
        COVID-related concern & 3 \\
        Monetary damage & 2 \\
        Sending to the wrong person & 2 \\ \bottomrule
    \end{tabulary}
\end{table}

We asked 
participants to describe potential risks associated with 
transmitting sensitive data generally, not in the context of a specific scenario.
Participants overwhelmingly referred to risks to the data at rest,
after transmission, rather than risks in transit, as shown in
Table~\ref{tab:risks_abbr}. 

Some examples include P5, who says, ``A facility or institution misplacing, losing, or selling my information to a 3rd party can be worrisome.'' P30 worries about ``Not knowing if the information will be kept safe.'' and P9 notes that ``the place I give these documents stores or disposes of them'', presumably indicating that if this storage and disposal is done improperly, their data will be at risk.
It is notable that participants almost always identified risks at the 
recipient, rather than risks involving themselves.

In general, participants did not provide many specific details when asked to identify risks. 
We used the broad categories that they identified as well as concerns they raised about the data at rest to inform our design for \surveyTwo{}. 
This allowed us to collect more details on the perceived risks of sending sensitive documents.

\paragraph{\surveyTwo{}}

\begin{figure}[t]
\includegraphics[width=\columnwidth]{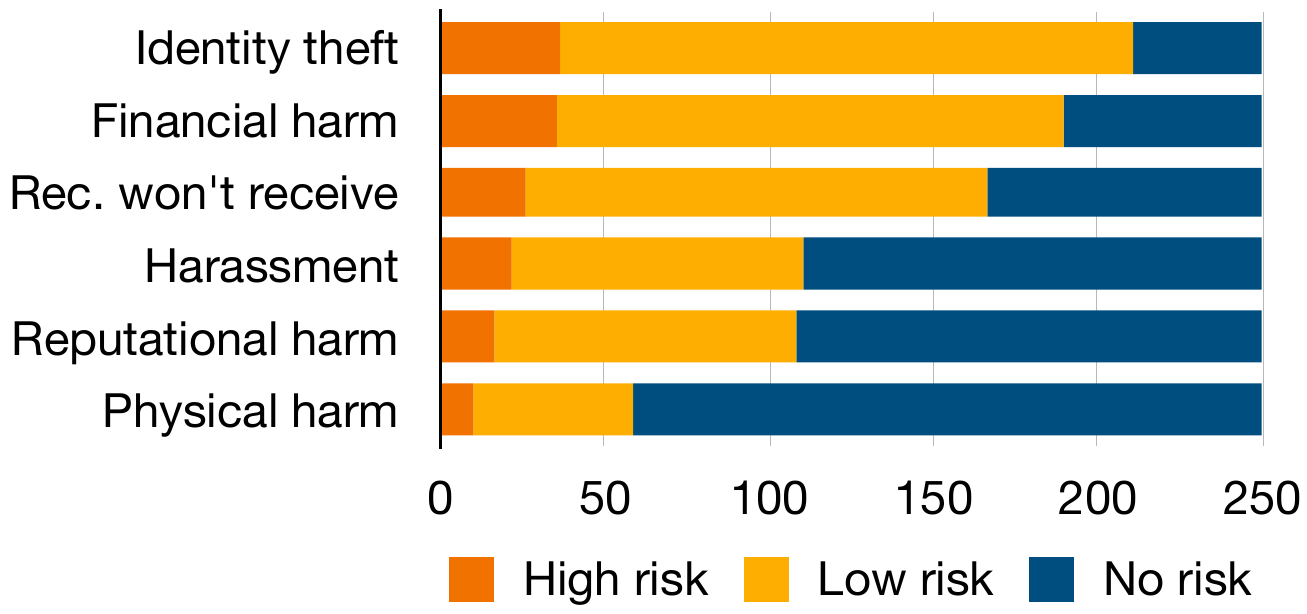}
\vspace{-.2in}
\caption{ Reported risk levels of various types of harm across all methods (\surveyTwo{})}
\label{fig:risks_overall_2}
\end{figure}

\begin{figure}[t]
\includegraphics[width=\columnwidth]{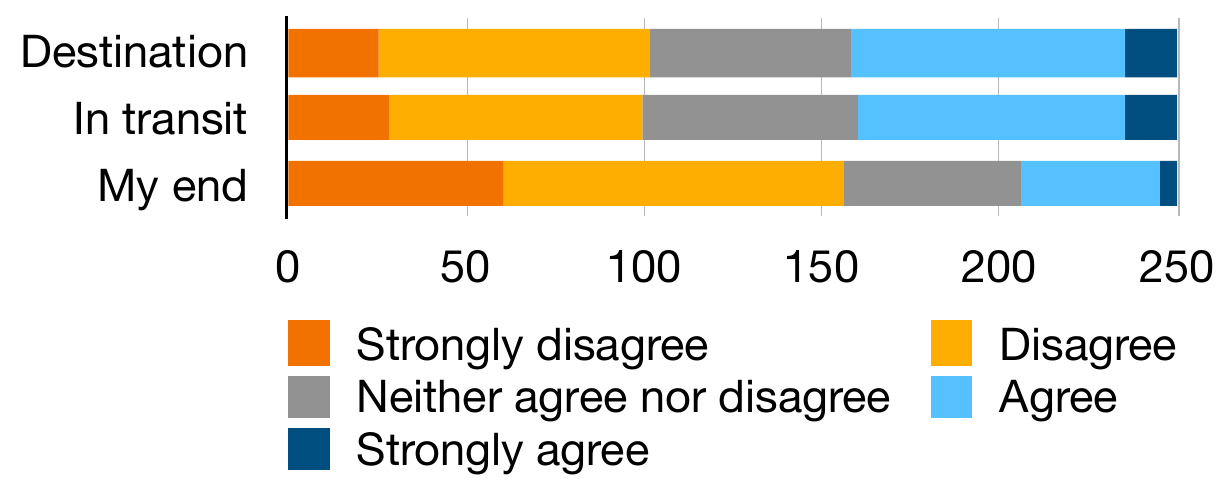}
\vspace{-.2in}
\caption{Where the risk is when sending sensitive information across all methods (\surveyTwo{})}
\label{fig:where_risk_overall_2}
\end{figure}

\begin{figure}[t]
\includegraphics[width=\columnwidth]{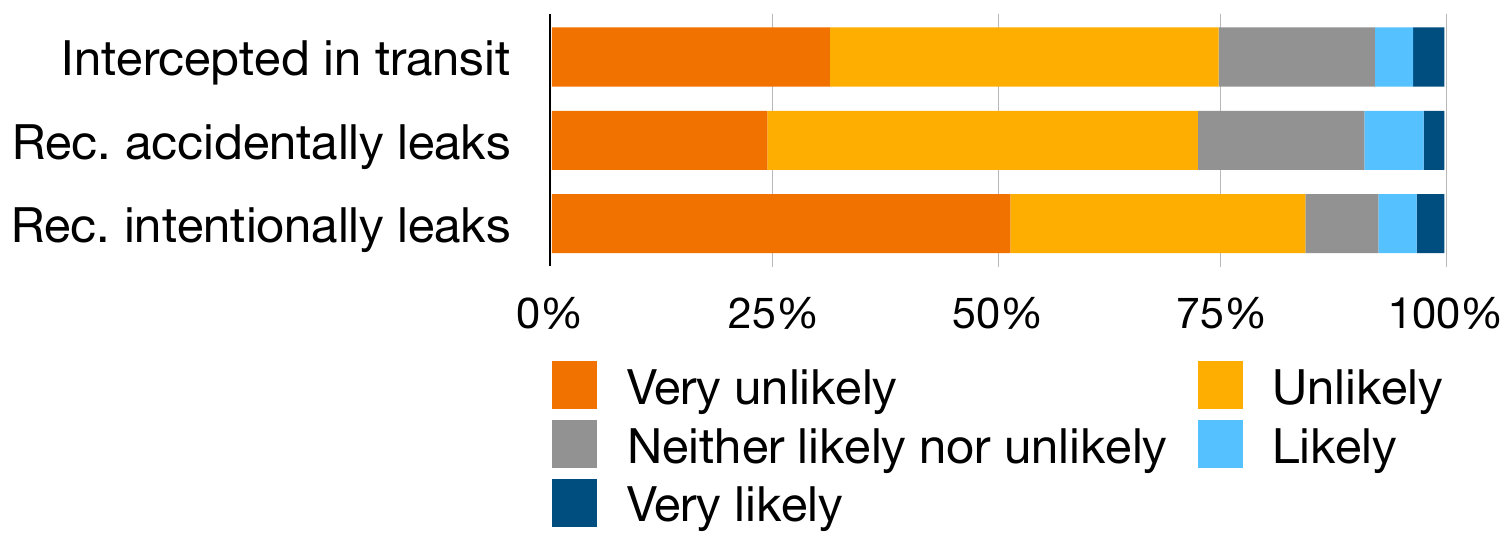}
\vspace{-.2in}
\caption{Likelihood of data to be leaked in various ways across all methods (\surveyTwo{})}
\label{fig:how_likely_2}
\end{figure}

We asked participants Likert-type questions based 
on the risks reported in \surveyOne{}, as well as additional risks 
that we considered interesting or important.
First, we asked whether --- for the specific incident we had 
asked them to recall --- they believed there was high risk, 
low risk, or no risk for a set of consequences, such as financial harm, 
reputational harm, or harassment. Participants overwhelmingly 
reported no or low risk (Figure~\ref{fig:risks_overall_2}). 
Slightly more risk was reported for identity theft and 
financial harm than for other concerns, which aligns  
with the prevalence of sending financial information and 
SSNs. Post-hoc, pairwise MWU comparisons (see Section~\ref{sec:methods_survey_2}) indicate that participants found identity theft and financial harm to be significantly more likely to be harmful. (Full details are given in Table~\ref{app:kw:risk_severity} in Appendix~\ref{app:figstables}.)

Figure~\ref{fig:how_likely_2} reports participant perception of how likely data might be to be leaked when transmitting sensitive documents based on statements derived from free responses in \surveyOne{}.
``Intercepted in transit'' refers to data being intercepted between the source and destination, while the other two statements refer to the recipient revealing sensitive information to a third party, either deliberately or by accident.
Most of these scenarios are viewed as ``very unlikely'' or ``unlikely''.
Post-hoc, pairwise MWU comparisons found that participants considered the recipient intentionally leaking their information to be significantly less likely than the data being intercepted or leaked by accident. (Full details are given in Table~\ref{app:kw:how_leaked} in Appendix~\ref{app:figstables}.)

We also investigated where in the transmission process participants view risk. Figure~\ref{fig:where_risk_overall_2} demonstrates 
that in general, participants identify more risk at the destination 
and in transit than in their own stewardship. Similar post-hoc pairwise 
tests (Table~\ref{tab:wilcoxon}) confirm that 
risk is perceived to be significantly greater 
at the destination and in transit than at the participant.

\begin{table}[t]
    \centering
    \small
    \caption{Post-hoc comparisons of risks using pairwise Mann-Whitney U-test with Holm-\v{S}id\'{a}k correction. (Omnibus Kruskal-Wallace test significant, $H=44.23$, $p<0.001$)}
    \label{tab:wilcoxon}
    \smallskip
    \begin{tabular}{lrr}
        \toprule
        \textbf{Comparison} & \textbf{\emph{p}} \\
        \midrule
        My end vs. in transit & < 0.001* \\ %
        My end vs. the destination & < 0.001* \\
        In transit vs. the destination & 0.922\phantom{*} \\ %

        \bottomrule
    \end{tabular}

\end{table}

\paragraph{Key findings for RQ3}
Participants are primarily concerned about financial harm and identity theft rather than risks of harassment or reputational damage. 
When unprompted, participants are most concerned with what happens to the data at its destination; after prompting, they express concern about risk in transit but do not identify risk at the sender (themselves).

 \section{Discussion}
In two surveys, we explored how users transmit sensitive information when required to do so, their privacy satisfaction with their transmission methods, and the risks associated with these interactions.
In \surveyOne{}, we presented participants with three scenarios and asked them to qualitatively describe a transmission method they used or imagined they would use. Building on those responses, in \surveyTwo{}, participants were randomly assigned
to a transmission method --- among eight methods identified in \surveyOne{} --- that they
had previously used successfully. We then asked them to recall a specific instance of
using that method to send sensitive information and answer closed-item questions about
their privacy satisfaction, convenience, and risk factors. These questions were also
derived from our qualitative coding of the results of \surveyOne{}.
In both surveys, participants generally described high satisfaction with both the convenience and privacy of their transmission methods and primarily described low risks. In most cases, the majority of participants indicated they would use the same transmission method again.

In this section, we explore larger themes and implications of the results, particularly around how participants see risks in transmitting sensitive information and choose a transmission method, as well as design implications and recommendations.

\paragraph{Familiarity and use are different}
In \surveyOne{}, email, online forms, and taking the documents in person were the most common methods participants suggested without prompting.
This raised an important question: do participants deliberately choose these methods over others, or have they simply not heard of alternatives?

The results from \surveyTwo{} answer this question: Large majorities of
participants had heard of all of the transmission methods.
Further, participants tended to be satisfied with their transmission methods,
regardless of whether they were prescribed by the recipient.
This suggests that targeting recipients of sensitive data --- like tax professionals and school personnel --- for education and advocacy could
have a positive impact on the security and privacy of these transmissions.

Additionally, there may be significant benefits to actively encouraging the use of tools that \emph{already} exist to perform this task, such as document-sharing services.
Our results suggest that participants know these options exist but simply do not choose to use them often. However, participants who discussed these methods did generally
believe they are secure and were usually satisfied with them. Making these services
more salient --- perhaps by evangelizing them to common document recipients ---
could provide useful benefits.
We also scoped this study to sending documents as a discrete transaction rather than continuous collaboration, which is an interesting but separate use case for which document sharing services might be more commonly used.
Continuous collaboration
on sensitive documents is a promising avenue for potential future work.

\paragraph{Only some information is considered sensitive}
When prompted in \surveyTwo{} to recall a situation where they sent sensitive
information, participants overwhelmingly selected financial information and
Social Security numbers. Very few participants' exemplar scenarios included
other identifying or secret information (contact details, passwords, etc.), suggesting that this information is considered less
sensitive, or is at least less likely to be top-of-mind when imagining
sensitive data.
This aligns with prior work that finds people have different standards for what information is considered sensitive~\cite{ruoti2017weighing,wash2015too,woodruff2014would}.
It also illuminates a potential gap, in which people may be transmitting sensitive
information without realizing the need to take precautions.
Future work could more directly examine what triggers people to recognize
``sensitive'' situations and consider communications privacy.

\paragraph{Risks at the endpoints}
In \surveyOne{}, participants primarily focused on data leaks at the
recipient, rather than in transit or at the source (e.g., from the user's
email account). In \surveyTwo{} --- when prompted with specific choices --- participants identified
risks in transit with similar frequency to risks at the destination,
but risks at the sender remained unrecognized.
This raises two key points.

First, the usable privacy community has primarily focused on risks to data in transit. This includes studies of secure email and messaging adoption~\cite{abu2017obstacles,abu2018exploring,de2016expert,ruoti2013confused}, as well as challenges in conveying proper and secure transmission, particularly with respect to certificate warnings and phishing~\cite{schechter2007emperor,sunshine2009crying,bravo2013your,akhawe2013alice,petelka2019put,felt2014experimenting}.
This aligns with our finding that people were not entirely confident their transmission method was secure.
While risks in transit are clearly important, our results suggest more attention should also be paid to risks at the endpoints, including, e.g., how to convey meaningful assurance that data is being handled properly at the destination.

Second, this finding accords with prior work showing that retrospective risks related to sensitive data left in one's own possession (often after sending it to someone) are opaque to end users~\cite{snyder2013cloudsweeper,clark2015saw,khan2018forgotten}.
Further work is needed to develop tools for both senders and recipients to clean up
no-longer-needed data, and educational interventions that teach about secure
communications should make sure to point out potential risks at the source as
well as in transit and at the destination.

\paragraph{Design implications and recommendations}
Our findings suggest opportunities to improve the design of current transmission methods for sensitive content. In particular, methods should take into account both endpoints, not just security in transit, and the security mechanisms should be as transparent as possible to the user to reduce overhead of using the method. Below, we outline where these results can be applied to certain application spaces.

\subparagraph{Document sharing services}
As mentioned above, document sharing services like Google Drive or Dropbox may provide a convenient and secure method to send sensitive documents.
Further research is needed into why these services are used less frequently and what can be done to increase their use.
Our participants who had tried them tend to believe they are secure and convenient, but many
have not tried.

\subparagraph{Confidential mode}
One attempt to improve transmission of sensitive data is Gmail's existing \emph{confidential mode}.\footnote{\url{https://support.google.com/a/answer/7684332} (viewed Feb 3, 2020)}
This service encodes an email as an image so the content cannot be printed and will be automatically deleted at a later time, after which the recipient will not be able to view the content.
While we know of no direct research on the efficacy of this method, the approach of interceding during the email process has promise, as both surveys and prior work~\cite{monson2018usability,bondarenko2015docshand} suggest that email is a common
approach for sending sensitive content, particularly when an alternative is unknown to either party.

Researchers should examine how to best intercede with the user workflow when opting for email based transmission of sensitive data. The user could then be prompted to apply a better mechanism first. However, the design and frequency of these interventions need to be carefully considered as prior work~\cite{egelman2018warning,felt2012android} suggests that very frequent security warnings are likely to be ignored by users. Such interventions need to map to peoples' risk models to be most effective. If a user does not see (or understand) a risk, they are unlikely to make the right choice~\cite{wogalter2006warnings}.

\subparagraph{Secure message deletion} Our study and other recent work~\cite{monson2018usability} show that users are concerned about their data even after it arrives at the destination. One suggestion applicable to email is for senders to use short-lived encryption keys per message that can expire or be revoked~\cite{monson2018usability}, similar to popular chat applications such as WhatsApp and Signal~\cite{perrin2016doubleratchet}. While promising, this idea inherits significant key management challenges~\cite{ruoti2016we, sweikata2009usability, garfinkel2005johnny, bai2016inconvenient, koh.bellovin.ea:easy} related to the decentralized nature of email, and it remains unknown if users will simply copy or screenshot the emails outside the secure email system to retain access. This is a promising area of future work.

\subparagraph{No-effort privacy}
While interventional approaches, such as prompting a user to use confidential mode, are important, an even better approach would be to offer users a transparent way to send sensitive information. This is analogous to incorporating end-to-end encryption into already popular messaging tools. For example, when including a potentially sensitive attachment, the document could automatically be conveyed via a secure document-sharing service, then automatically retrieved at the destination. This would allow the workflow at both endpoints to continue unchanged. This could parallel existing processes in email services that partially or entirely automatically send large attachments via cloud storage links rather than directly via email. As part of providing this service, additional work might be needed to convey the additional privacy benefits; demonstrating when
communication is private has proven challenging in domains from web browsing to secure
messaging~\cite{felt2016rethinking,tan2017unicorns,schechter2007emperor}.

\subparagraph{Retrospective privacy}
Our results also confirm the previously identified need~\cite{khan2018forgotten} for retrospective privacy. Email, cloud storage,
and document sharing service providers could offer automated suggestions for deleting older content --- both sent and received --- and
automatic message expiration options~\cite{monson2018usability}.
Providers could offer an option to mark sensitive content when it is created, to allow for review and potential deletion in the future. This could be modeled on approaches that allow users to ``snooze'' an email for future action or nudge users to revisit content that has not been accessed in a while.
Elements like these could help users protect content at rest, even if it was not protected at transmission time.

 \section{Conclusion}
\label{sec:conc}

This paper reports on two surveys of users' experiences sending sensitive information: \surveyOne{} using common scenarios drawn from prior work as prompts
($n=60$, 180 total scenario instances) and \surveyTwo{} ($n=250$) asking more detailed questions based on results from \surveyOne{}.
We found that users most frequently expect to 
deliver documents in person or to use email; in 
reality, they typically use these methods as well 
as online portals or forms provided by institutional 
recipients. 
We also found that participants report high levels 
of satisfaction with the privacy and convenience 
of their existing methods, while recognizing that 
there are possible risks associated with transmitting 
this information, particularly risks of data leaking 
after being received at the destination. These 
results suggest new opportunities for tools and 
user interventions designed to make secure 
transmission of documents simpler and more 
transparent, and supporting retrospective privacy 
by nudging users to delete no-longer-needed content.

 \subsection*{Acknowledgements}
\label{sec:acks}

We gratefully acknowledge support from a UMIACS contract under the partnership between the University of Maryland and DoD. The views expressed are our own.

We’d also like to thank the reviewers for their insightful comments and feedback, as well as Kelsey Fulton, Omer Akgul, Nathan Reitinger, and the other members of the SP2 and GWUSEC labs for their help and support.

 \bibliographystyle{plain}
 \bibliography{sending}

\appendix

\section*{Appendix}

\section{Scenarios}
\label{app:scen}

The following are the nine scenarios we presented to the subjects, exactly as they were shown in the questionnaire.

\paragraph{Applying for a Mortgage}

You are applying for a mortgage so you can purchase a new home. You must send the bank the following information.

\begin{itemize}[noitemsep]
    \item Proof of income - W-2 forms and two most recent payroll stubs or other income information
    \item 60 days worth of bank statements
    \item Monthly debt payment information - car payments, student loan payments, credit card debt payments
    \item Rent payment for the past twelve months
    \item Divorce decree, if applicable
\end {itemize}

\paragraph{Sharing a Password}

A trusted friend needs access to an email account the two of you share. You need to send them the password to this account.

\paragraph{Background Check}

You are interested in doing some volunteer work, and the group you are working for has asked you to do a background check. You must send the volunteer group the following information.

\begin{itemize}[noitemsep]
    \item Full name
    \item Social Security Number
    \item Date of birth
    \item All addresses where you have lived in the past 5 years
    \item Names and contact information for two personal references
\end{itemize}

\paragraph{Applying for an Apartment}

You are applying to rent an apartment and are preparing your paperwork. You are required to send the landlord all of the following documents or information.

\begin{itemize}[noitemsep]
    \item Basic demographic information - name, email, phone number
    \item Emergency contacts
    \item Social Security Number
\end{itemize}

\paragraph{Opening a Checking Account}

You are opening a checking account at a new bank. The bank requires you to send them the following information.

\begin{itemize}[noitemsep]
    \item Social Security Number and date of birth of all account holders
    \item Phone number and email address
    \item Physical U.S. address (no post office boxes)
    \item Debit card or account information for funding your new account
\end{itemize}

\paragraph{Sharing a Password-Protected Document}

You have a password-protected PDF that is encrypted. You need to share both the PDF and the password to open it with a trusted friend.

\paragraph{New School}

Your child is starting at a new school, and you must send the school copies of the following documents about your child.

\begin{itemize}[noitemsep]
    \item Birth certificate
    \item Proof of custody/guardianship
    \item Proof of residency like one of the following: current property tax bill, current rental lease, current utility \item Immunization record
    \item Social Security card
\end{itemize}

\paragraph{Seeing a New Doctor}

You are going to see a new doctor for the first time. You are asked to send the new doctor’s office the following information.

\begin{itemize}[noitemsep]
    \item Current insurance information
    \item An image of your driver’s license or other valid photo ID
    \item A list of any medication you are currently taking
    \item Your health history
\end{itemize}

\paragraph{Sending Tax Documents}

You are getting ready to prepare your taxes and have hired a Certified Public Accountant (CPA). They ask you to send them the following information.

\begin{itemize}[noitemsep]
    \item A copy of your Social Security card
    \item All income-related tax documents - W-2, 1099, etc.
    \item All expense-related tax documents - 1098, rental expenses, etc.
\end{itemize}

\clearpage

\section{Additional Tables}
\label{app:figstables}

\begin{table}[h]
    \centering
    \small
\caption{Free responses to ``What other methods have you used?'' (\surveyTwo{}). Many participants repeated methods that were provided in the closed-answer question.}
\label{tab:other_methods_2}
\smallskip
    \begin{tabular}{l r}
    \toprule
    \textbf{Transmission Method} & \textbf{\#Part.} \\
        \midrule
        Online form or portal & 20 \\
        In person & 14 \\
        Physical mail & 11 \\
        Email & 9 \\
        Direct messaging & 8 \\
        Online (no further spec.) & 5 \\
        Courier service & 4 \\
        Email (mentions encryption) & 4 \\
        Fax & 3 \\
        Phone call & 3 \\
        Direct messaging (mentions encryption) & 2 \\
        Document sharing service & 2 \\
        Via flash drive & 2 \\
        Encryption (no further specification) & 1 \\
        Live chat support & 1 \\
        Used a VPN & 1 \\
        Via encrypted flash drive & 1 \\
        \bottomrule
    \end{tabular}
\end{table}

\begin{table}[h]
    \centering
    \small
\caption{Free responses to ``What kind of information were you sending: Other'' (\surveyTwo{})}
\label{tab:what_sending_other_2}
\smallskip
    \begin{tabular}{lr}
    \toprule
    \textbf{Data Type} & \textbf{\#Part.} \\
        \midrule
        Home address & 12 \\
        Identity documents & 12 \\
        Demographic details &  9 \\
        General personal information & 6 \\
        Financial information & 5 \\
        Contact information &  4 \\
        Login credentials & 3 \\
        Work documents &  3 \\
        Titles and deeds & 2 \\
        Insurance documents & 1 \\
        \bottomrule
    \end{tabular}
\end{table}

\begin{table}[h]
    \centering
    \small
\caption{Factor levels for inputs to ordinal logistic regression before model selection. Asterisks (*) indicate baselines.}
\label{tab:factor_levels}
\smallskip
\resizebox{\linewidth}{!}{
    \begin{tabular}{l l}
    \toprule
    \textbf{Factor} & \textbf{Levels} \\
        \midrule
        \textbf{Method} &  In person$^*$ \\
        Categorical &	Physical mail \\
        &	Direct messaging \\
        &	Online form or portal \\
        &	Phone call \\
        & 	Fax \\
        &	Email \\
        &	Document sharing service \\
        \midrule
        \textbf{What was being sent} & Financial info \\
        Binary for each option as & Social Sec. number \\
        participants could choose multiple. & Info about children \\
        Baseline for each was false & Info about health \\
        & Sexual or explicit content \\
        & Other (please specify) \\
        
        \midrule
        \textbf{Recipient} & An organization$^*$ \\
        Categorical &	A particular professional \\
        &	A friend, partner, or family member \\
        &	An employer or potential employer \\
        &	A landlord or potential landlord \\
        &	A gov't or gov't institution \\
        &	Other (please specify) \\
        \midrule
        \textbf{Who chose the method} & I suggested$^*$ \\
        Categorical &	Recipient suggested \\
        &	Recipient required \\
        &	Discussed jointly \\
        \midrule
        \textbf{Trust in recipient} & Don't trust$^*$ \\
        Binned from 5pt Likert & Trust \\
        \midrule
        \textbf{Agreement with statements} & Disagree$^*$ \\
        Each below binned from 5pt Likert & Agree \\
        \textit{I am tech-savvy} & \\
        \textit{The rec. is tech-savvy} & \\
        \textit{Risk is at my end} & \\
        \textit{Risk is at dest.} & \\
        \textit{Risk is at their end} & \\
        \midrule
        \textbf{Likelihood of leaks} & Unlikely$^*$ \\
        Each below binned from 5pt Likert & Likely \\
        \textit{Rec. will accidentally leak} & \\
        \textit{Rec. will intentionally leak} & \\
        \textit{Data intercepted in transit} & \\
        \midrule
        \end{tabular}}
\end{table}

\begin{table*}
\centering
\small
\caption{
Post-hoc comparisons of convenience satisfaction (\surveyTwo{}, Q13) across transmission methods using pairwise Mann-Whitney U-test with Holm-\v{S}id\'{a}k correction. (Omnibus Kruskal-Wallace test significant, $H=42.56$, $p<0.001$)}
\label{tab:appendix:kw-convenience}
\begin{tabular}{rccccccc}
\toprule
{ } & \textbf{In Person} & \textbf{Online Port.} & \textbf{Email} & \textbf{Physical Mail} & \textbf{Phone} & \textbf{Doc. Share} & \textbf{Fax} \\
\midrule
In Person & --- &   &   &   &   &   &   \\
Online Port. & \phantom{<}0.461\phantom{*} & --- &   &   &   &   &   \\
Email & \phantom{<}0.630\phantom{*} & \phantom{<}0.952\phantom{*} & --- &   &   &   &    \\
Physical Mail & \phantom{<}0.024* & < 0.001* & < 0.001* & --- &   &   &   \\
Phone & \phantom{<}0.909\phantom{*} & \phantom{<}0.054\phantom{*} & \phantom{<}0.101\phantom{*} & \phantom{<}0.089\phantom{*} & --- &   &   \\
Doc. Share & \phantom{<}0.909\phantom{*} & \phantom{<}0.884\phantom{*} & \phantom{<}0.909\phantom{*} & \phantom{<}0.006* & \phantom{<}0.630\phantom{*} & --- &   \\
Fax & \phantom{<}0.461\phantom{*} & \phantom{<}0.012* & \phantom{<}0.020* & \phantom{<}0.884\phantom{*} & \phantom{<}0.785\phantom{*} & \phantom{<}0.149\phantom{*} & ---   \\
Direct Mesg. & \phantom{<}0.909\phantom{*} & \phantom{<}0.285\phantom{*} & \phantom{<}0.384\phantom{*} & \phantom{<}0.362\phantom{*} & \phantom{<}0.964\phantom{*} & \phantom{<}0.832\phantom{*} & \phantom{<}0.887\phantom{*} \\
\bottomrule
\end{tabular}
\end{table*}

\begin{table*}[t]
\centering
\small
\caption{Post-hoc comparisons of likelihood to reuse a given transmission method (\surveyTwo{}, Q16) using pairwise Mann-Whitney U-test with Holm-\v{S}id\'{a}k correction. (Omnibus Kruskal-Wallace test  significant, $H=38.53$, $p<0.001$)
}
\label{tab:appendix:kw-woulduse}
\begin{tabular}{rccccccc}
\toprule
{ } & \textbf{In person} & \textbf{Online Port.} & \textbf{Email} & \textbf{Physical Mail} & \textbf{Phone}  & \textbf{Doc.\ Share} & \textbf{Fax} \\
\midrule
In Person & --- &   &   &   &   &   &  \\
Online Port. & \phantom{<}0.989\phantom{*} & --- &   &   &   &   &   \\
Email & \phantom{<}0.067\phantom{*} & \phantom{<}0.032* & --- &   &   &   &   \\
Physical Mail & \phantom{<}0.001* & < 0.001* & \phantom{<}0.950\phantom{*} & --- &   &   &   \\
Phone  & \phantom{<}0.105\phantom{*} & \phantom{<}0.043* & \phantom{<}0.977\phantom{*} & \phantom{<}0.453\phantom{*} & --- &   &   \\
Doc. Share & \phantom{<}0.965\phantom{*} & \phantom{<}0.955\phantom{*} & \phantom{<}0.676\phantom{*} & \phantom{<}0.078\phantom{*} & \phantom{<}0.864\phantom{*} & --- &   \\
Fax & \phantom{<}0.018* & \phantom{<}0.006* & \phantom{<}0.982\phantom{*} & \phantom{<}0.977\phantom{*} & \phantom{<}0.925\phantom{*} & \phantom{<}0.365\phantom{*} & --- \\
Direct Mesg. & \phantom{<}0.081\phantom{*} & \phantom{<}0.041* & \phantom{<}0.989\phantom{*} & \phantom{<}0.960\phantom{*} & \phantom{<}0.971\phantom{*} & \phantom{<}0.676\phantom{*} & \phantom{<}0.986\phantom{*} \\
\bottomrule
\end{tabular}
\end{table*}

\begin{table*}[t]
\centering
\small
\caption{Post-hoc comparisons of severity of risks of the participant's transmission method (\surveyTwo{}, Q18) using pairwise Mann-Whitney U-test with Holm-\v{S}id\'{a}k correction. (Omnibus Kruskal-Wallace test significant, $H=249.69$, $p<0.001$}
\label{app:kw:risk_severity}

\begin{tabular}{rccccc}
\toprule
{ } & \textbf{Harassment} & \textbf{Identity theft} & \textbf{Financial} & \textbf{Physical} & \textbf{Reputational} \\
\midrule
Harassment & --- &   &   &   &  \\
Identity theft & < 0.001* & --- &   &   &  \\
Financial & < 0.001* & \phantom{<}0.158\phantom{*} & --- &   &  \\
Physical & < 0.001* & < 0.001* & < 0.001* & --- &  \\
Reputational  & \phantom{<}0.723\phantom{*} & < 0.001* & < 0.001* & < 0.001* & --- \\
Not Received & < 0.001* & < 0.001* & \phantom{<}0.045* & < 0.001* & < 0.001* \\
\bottomrule
\end{tabular}
\end{table*}

\begin{table*}[t]
\centering
\small
\caption{Post-hoc comparisons how likely sensitive data is to be leaked (\surveyTwo{}, Q14) using pairwise Mann-Whitney U-test with Holm-\v{S}id\'{a}k correction. (Omnibus Kruskal-Wallace test was  significant, $H=32.31$, $p<0.001$)}
\label{app:kw:how_leaked}
\begin{tabular}{rcc}
\toprule
{ } & \textbf{Intercepted in transit} & \textbf{Recipient leaks by accident} \\
\midrule
Intercepted in transit & --- & \\
Recipient leaks by accident & \phantom{<}0.259\phantom{*} & ---\\
Recipient leaks on purpose & < 0.001* & < 0.001* \\
\bottomrule
\end{tabular}
\end{table*}

\processifversion{arxiv}{

\begin{table*}[!ht]
  \caption{Transmission modes reported by participants in Survey 1, broken down by
  scenario as well as into real and imagined experiences. Counts indicate the
  total number of participants; for context, counts are also given for individual
  codes. Participant responses sometimes included more than one transmission method per instance.}
  \label{tab:scene}
  \centering
  \small
  \vspace{+.1in}

    \begin{tabular}{p{4.5cm}clcl}
    \toprule
      \multicolumn{3}{c}{\hspace{+1cm} \bf { \quad  \quad  Real \quad  \quad }} & \multicolumn{2}{c}{\hspace{-2.2cm}  \bf {\quad   \quad Imag. \quad  \ \,}}\\
      \cmidrule(r){2-3} \cmidrule(l){4-5}
      \multicolumn{1}{l}{\makecell[l]{\textbf{Scenario}}} & \makecell[c]{\textbf{Count}} & \makecell[l]{\textbf{Modes}} & \makecell[l]{\textbf{Count}}  & \makecell[l]{\textbf{Modes}} \\ \midrule
\\
        Renting an apartment & 15 & \makecell[l]{In person (11); online form (7); \\email (4); physical mail (2); \\secure online (1); fax (1); \\online (1); other (3)} & 5 & \makecell[l]{Email (3); in person (1); \\online (1)} \\ \\ %

        Background check & 11 & \makecell[l]{In person (6);  email (5); \\online form (4); direct message (1); \\ online (1); physical mail (1); \\ video call (1); other (5) } & 9 & \makecell[l]{In person (4);\\ direct message (1);\\ other (4)} \\\\ %

        Applying for a mortgage & 7 & \makecell[l]{Online form (3); in person (2); \\document sharing (1); \\email (1); online (1);\\ other (2)} & 13 & \makecell[l]{In person (8); email (5); \\ online form(1); \\fax (1); online (1)} \\\\ %

        Creating a new checking account & 18 & \makecell[l]{Online form (10); email (3); \\phone call (2); secure online (2); \\fax (1); online (1);\\ physical mail (1); other (3)}  & 2 & \makecell[l]{In person (1); \\other (1)} \\ \\ %

        Visiting a new doctor & 15 & \makecell[l]{In person (13); online form (4); \\ email (2); online (2); \\direct message (1); physical mail (1); \\ fax (1); other (4)} & 5 & \makecell[l]{Physical mail (1); \\in person (1); fax (1); \\email (1); other (1)} \\ \\ %

        Starting a child at a new school & 7 & \makecell[l]{In person (4); email (3); \\online form (1); physical mail (1); \\secure online (1); other (4)} & 13 & \makecell[l]{In person (9); email (2); \\fax (1); online form (1);\\ other (1)} \\\\ %

        Sharing a password to an email account & 11 & \makecell[l]{Direct messaging (7); \\phone call (4); in person (3); \\email (1); other (1)} & 9 & \makecell[l]{Phone call (4); \\direct messaging (3); \\in person (2); secure online (1)} \\ \\ %

        Sharing a password-protected PDF & 7 & \makecell[l]{Direct messaging (5); \\email (5); phone call (3); \\doc.\ sharing service (1); \\other (1)} & 13 & \makecell[l]{Email (8); phone call (6); \\direct messaging (4); \\doc. sharing service (1); \\in person (1); other (2)} \\ \\ %

        Sharing financial information with a CPA & 6 & \makecell[l]{In person (4); online form (1); \\physical mail (1); \\secure online (1); other (4)} & 14 & \makecell[l]{Email (9); in person (4); \\fax (2); other (1)} \\

        \bottomrule
    \end{tabular}
\end{table*}

\clearpage}

\processifversion{usenix}{
\section{Extended Appendices}
\label{app:extend}

An extended version of the paper including the full text of each survey and the qualitative codebook for \surveyOne{} can be found at \href{https://google.com}{ARXIVSOMETHINGSOMETHING}.}

\processifversion{arxiv}{
\section{Survey 1}
\label{app:survey1}

Note: The questions directly relating to COVID-19 were only presented in the second round of data collection.

\subsection{Scenarios}

In this survey, you will be presented with three scenarios, and asked a series of questions about each one. Please read each scenario carefully and answer the questions as accurately as possible.

Additionally, please note that there is no correct answer for many of these questions - please answer them as honestly as you are able.

Please read the following scenario carefully before proceeding. The next set of questions will be asking about this scenario - the scenario is printed in its entirety on every page.

You are getting ready to prepare your taxes and have hired a Certified Public Accountant (CPA). They ask you to send them the following information.

\begin{itemize}[noitemsep]
    \item A copy of your Social Security card
    \item All income-related tax documents - W-2, 1099, etc.
    \item All expense-related tax documents - 1098, rental expenses, etc.
\end{itemize}

\begin{enumerate}
    \item Even if the details are slightly different, have you ever been in a situation like this one? [Yes, No, Not sure]
\end{enumerate}

The following questions were presented if the participant had encountered the scenario before.
\begin{enumerate}
    \item Think back to when you did this. How did you provide the required information? [free text]
    \item When did the scenario you are thinking of happen? [Before social distancing due to COVID-19, After social distancing due to COVID-19, Both before and after social distancing due to COVID-19, Not sure, Prefer not to say]
    \item if Q2 = After or Both: How did social distancing affect the method you chose? [free text]
    \item if Q2 = Before, Not sure, or Prefer not to say: Imagine you had to do this while social distancing. How would your method change, if at all? [free text]
\end{enumerate}

The following questions were presented if the participant had not encountered the scenario before.
\begin{enumerate}
    \item \textbf{Please imagine} you were in this situation. How would you provide the required information? [free text]
    \item Would your method be impacted by social distancing due to COVID-19? [Yes, No, Not sure, Prefer not to say]
    \item Please explain your answer to ``Would your method be impacted by social distancing due to COVID-19?'' [free text]
\end{enumerate}

The remaining questions were presented regardless of whether the participant had encountered the scenario, except for Q7 in the following block, which was only presented to people who had encountered the scenario.
\begin{enumerate}
    \item How satisfied are you with the method(s) you described for providing this information? [Very unsatisfied, Unsatisfied, Neither satisfied nor unsatisfied, Satisfied, Very satisfied]
    \item Can you provide some more details regarding your satisfaction level with the method(s) you used? [free text]
    \item How satisfied are you with the \textbf{security and privacy} of these method(s) for providing the information? [Very unsatisfied, Unsatisfied, Neither satisfied nor unsatisfied, Satisfied, Very satisfied]
    \item Can you provide some more details regarding your satisfaction level in terms of \textbf{security and privacy}? [free text]
    \item How satisfied are you with the \textbf{convenience} of these method(s) for providing the information? [Very unsatisfied, Unsatisfied, Neither satisfied nor unsatisfied, Satisfied, Very satisfied]
    \item Can you provide some more details regarding your satisfaction level in terms of \textbf{convenience}? [free text]
    \item Can you think of a better way to provide this information? If so, please describe it below. [free text]
\end{enumerate}

\subsection{Risks, Mitigations and Demographics}

Note: the following questions were only presented once, after the above questions had been answered about the three randomly selected scenarios.

\begin{enumerate}
    \item All three of the scenarios that you looked at dealt with providing required documents or information. There are sometimes risks associated with providing required documents like these. What, if anything, concerns you? Please provide two examples. [two free text boxes]
    \item Sometimes it is possible to take precautions to reduce the risks you just identified. Can you think of any precautions a person could take to alleviate these risks? Please provide an answer for each example you gave previously. [two free text boxes]
    \item Have you ever taken precautions like these? [Yes, No, Not sure]
    \item Please explain your answer to ``Have you ever taken precautions like these?'' [free text]
    \item Do you have a computer science background? This means working in or holding a degree in computer science or information technology. [Yes, No, Not sure, Prefer not to say]
    \item Please specify the range which most closely matches your total, pre-tax, household income in 2019. [Less than \$10,000, Between \$10,000 and \$14,999, Between \$15,000 and \$24,999, Between \$25,000 and \$34,999, Between \$35,000 and \$49,999, Between \$50,000 and \$74,999, Between \$75,000 and \$99,999, Between \$100,000 and \$149,999, Between \$150,000 and \$199,999, Over \$200,000, Prefer not to say]
    \item Have you ever held a government security clearance? [Yes, No, Not sure, Prefer not to say]
    \item Have you ever handled any of the following information in a professional context? Please check all that apply. [Information covered by the Health Insurance Portability and Accountability Act (HIPAA), Information covered by the Family Educational Rights and Privacy Act (FERPA), Social Security numbers, Credit card information]
    \item Is there any feedback on our survey or additional information you'd like to provide to help us understand your responses or improve the survey? [free text]
\end{enumerate}

\section{Codebook}
\label{app:codebook}

\begin{flushleft}
   \textbf{Methods}
   \begin{enumerate}
     \item Document sharing service (Dropbox, Google Drive, Box, etc.)
     \item Physical mail (Fedex, USPS, etc.)
     \begin{enumerate}
       \item Certified mail
     \end{enumerate}
     \item In person
        \begin{enumerate}
            \item By speaking
            \item On paper
            \item Unspecified
        \end{enumerate}
    \item Secure sending online (no further specification)
    \item Online unspecified
    \item Online form/portal (provided by recipient)
    \item Phone call
    \item Video call
    \item Fax
    \item Email
        \begin{enumerate}
            \item Encrypted
            \item Unencrypted
            \item Unencrypted with encrypted attachment
            \item Unspecified
        \end{enumerate}
    \item Direct messaging
        \begin{enumerate}
            \item Secure app (WhatsApp, Signal, etc.)
            \begin{enumerate}
                \item Mentions security
                \item Does not mention security
            \end{enumerate}
            \item Insecure app (Facebook Messenger, Snapchat, etc.)
            \item SMS
            \item Unspecified
        \end{enumerate}
    \item Can't think of a better way
    \item Other (including answers that are not methods)
   \end{enumerate}

   \textbf{General Satisfaction}
   \begin{enumerate}
       \item The method was secure
       \item The method was insecure
       \item The method was convenient
       \item The method was inconvenient
       \item The intended recipient received the information
       \item Concern related to COVID-19
       \item No better option
       \item Would prefer another method
       \item Other
   \end{enumerate}

    \textbf{Privacy Satisfaction}
        \begin{enumerate}
            \item Satisfied with the privacy of my method
            \begin{enumerate}
                \item At rest
                \begin{enumerate}
                    \item Recipient would keep my information safe (recipient has good security procedures)
                    \item I can keep my information safe (delete local copies, give recipient instructions to ensure safety)
                \end{enumerate}
                \item In transit
                \begin{enumerate}
                    \item Method of sending is secure
                    \item Information would be received by the intended recipient
                \end{enumerate}
            \end{enumerate}
            \item Unsatisfied with the privacy of my method
                \begin{enumerate}
                    \item At rest
                    \begin{enumerate}
                            \item Recipient will unintentionally reveal information (hack, breach, etc.)
                            \item Recipient will intentionally reveal information (sell, malicious intent, etc.)
                    \end{enumerate}
                    \item In transit
                        \begin{enumerate}
                            \item Method of sending is insecure (information could be intercepted in transit)
                        \end{enumerate}
                    \end{enumerate}
            \item Concern related to COVID-19
            \item Unsure about the security of the method
            \item Other/unrelated
        \end{enumerate}

    \textbf{Convenience Satisfaction}
    \begin{enumerate}
        \item The method was convenient (no further specification)
        \begin{enumerate}
            \item Use codes from Methods section if specific method is provided
        \end{enumerate}
        \item The method was neither convenient nor inconvenient
        \item The method was inconvenient (no further specification)
        \begin{enumerate}
            \item Use codes from Methods section if specific method is provided
        \end{enumerate}
        \item Gathering the required information is inconvenient
        \item No better option
        \item Concern related to COVID-19
        \item Other
    \end{enumerate}

    \textbf{Risks}
    \begin{enumerate}
        \item Identity theft
        \item Monetary damage
        \item Damage to reputation
        \item Someone would acquire the information for malicious purposes
        \item The data at rest is at risk
        \item The data could be lost/misplaced
        \item Someone (corporation, government) could read my emails
        \item The data in transit is at risk (including interceptions, other unspecified situations, etc.)
        \begin{enumerate}
            \item Technical failures
        \end{enumerate}
        \item Concern related to COVID-19
        \item Accidentally sending information to the wrong person
        \item Other
    \end{enumerate}

    \textbf{Mitigations}
    \begin{enumerate}
        \item Send the information securely (no further specification)
        \item Send the information digitally (no further specification)
            \item Change my password
            \item Have a password
            \item Use a VPN
            \item Use encryption (nonspecific)
            \item Keep sensitive physical documents safe
            \begin{enumerate}
                \item Store in a bag that zips
                \item Properly dispose of physical documents
            \end{enumerate}
            \item Only give sensitive information to trusted recipients
            \begin{enumerate}
                \item The recipient should have good/secure internal processes for handling my information
                \item Verify recipient followed the my instructions
                \item Verify recipient needs the information for a valid reason
                \item Give the recipient limited information
            \end{enumerate}
            \item Deletion
            \begin{enumerate}
                \item Ask recipient to delete my data
                \item Delete my local copy of the data
            \end{enumerate}
            \item Properly identify myself to the recipient
            \item Send sensitive information promptly
            \item Be aware of leaked information/documents (no further specification)
                \begin{enumerate}
                    \item Watch your credit score
                \end{enumerate}
            \item Only give out sensitive information when necessary
            \item Specified method of sending
            \begin{enumerate}
                \item Use codes from Methods section if specific method is provided
            \end{enumerate}
            \item Donate money to cybersecurity organizations
            \item Store information electronically
            \item Use a safe third party service
            \item Avoid delivering in person
            \item Avoid delivering online
            \item Record the transaction
            \item Concern related to COVID-19
            \item I can't think of another way to mitigate these risks
            \item Other

        \end{enumerate}
    \textbf{Precautions Taken}
    \begin{enumerate}
       \item I haven't taken these precautions (no further specification)
       \begin{enumerate}
           \item I am unconcerned about policy
           \item These precautions are a hassle/too much work
           \item I don't know what these precautions would be
           \item I have not been in these situations (referring to the scenarios in the survey)
           \item I can't take these precautions
           \item Taking these precautions would not be helpful
       \end{enumerate}
    \item I try to be secure/cautious (no further specification)
        \begin{enumerate}
            \item Use a VPN
            \item Use physical security (using a bag, safe, briefcase, etc.)
            \item Use an in person method
            \begin{enumerate}
                \item On paper
            \end{enumerate}
        \item Encrypted server/document storage system
        \item Email
            \begin{enumerate}
                \item Gmail
                \item Encrypted
                \item Unspecified
            \end{enumerate}
        \item Verify recipient identity
        \item Use password protection
        \end{enumerate}
        \item I avoid certain methods
        \begin{enumerate}
            \item I avoid using phone calls as my method of sending
        \item I avoid putting passwords in an unsafe location
        \end{enumerate}
        \item My network/home/office is secure
        \item Recipient-related
        \begin{enumerate}
            \item I asked the recipient to delete my information
        \item I give my information to trusted individuals / I don't give my information to non-trusted individuals
        \end{enumerate}
        \item Habits
        \begin{enumerate}
        \item I have a habit of keeping my information well-organized
        \item I have a habit of changing my passwords
        \item I have a habit of checking my credit regularly
        \end{enumerate}
        \item Precaution related to COVID-19
        \item Other/Unsure
    \end{enumerate}

    \textbf{Imagine Situation and Social Distancing (COVID-19 Question)}
    \begin{enumerate}
        \item It would not change
        \item I would use/prefer to use a different method of sending
        \begin{enumerate}
            \item Use codes from Methods section if specific code is provided
        \end{enumerate}
        \item I would take precautions related to COVID-19 (wearing a face mask, not making contact, etc.)
        \item I would do it in person once COVID-19 is not a concern
    \end{enumerate}

    \textbf{Imagine Social Distancing (COVID-19 Question)}
    \begin{enumerate}
        \item I would expose myself or others to risks relating to COVID-19
        \item I would take precautions specific to COVID-19, but use an in-person method
        \item My method would not be impacted
        \item I would use a different method (no further specification)
        \begin{enumerate}
            \item Use codes from Methods section if specific code is provided
            \item Would use a third-party service
        \end{enumerate}
        \item I can't use my preferred method (no replacement specified)
    \end{enumerate}
\end{flushleft}

\section{Survey 2}
\label{app:survey2}
\begin{enumerate}
    \item We often have to send sensitive information to another person, even when we might not know them. Some examples include sending tax information to a Certified Public Accountant, providing personal information to apply for an apartment, or setting up a new checking account at a new bank. Which of the following methods have you used to send information like this? For each, choose one of the listed options. [Have successfully used it at least once, have attempted to use but was unsuccessful, aware of it but never tried, am not aware of it]
    \begin{enumerate}
        \item Emailed the information.
        \item Sent the information via direct messaging (SMS, texting, WhatsApp, Facebook Messenger etc).
        \item Provided the information in person.
        \item Used an online form or portal to submit the information.
        \item Used a document sharing service (Google Drive, Dropbox, Box etc) to share the information.
        \item Provided the information over the phone.
        \item Faxed the information.
        \item Physically mailed the information.
    \end{enumerate}

    \item Please list any other methods you have used to send sensitive information. [free text]

    [\emph{We will show the participants the remainder of this survey one time by randomly selecting one of the methods they indicate they have used. The exact number will depend on the length of time the survey takes which we will determine in the pilot process. [METHOD] will be substituted by the relevant method.}]

    \item Please think about one specific time when you sent sensitive information via [METHOD]. For the rest of this survey, we are going to ask you questions about this specific time you sent sensitive information.
    
    Please briefly describe this one specific time in the following box. [free text]

    \item In the scenario you described, what kind of information were you sending via [METHOD]? Please select all that apply.
    \begin{enumerate}
        \item Financial information
        \item Social security number
        \item Information about children
        \item Health or wellness information
        \item Sexual / explicit content
        \item Other (please specify)
    \end{enumerate}
    \item Why were you sending this information? [free text]
    \item When did this scenario take place?
    \begin{enumerate}
        \item Within the last week
        \item Within the last month
        \item Within the last year
        \item More than a year ago
    \end{enumerate}
    \item Who were you sending the information to via [METHOD]? [check all that apply]
    \begin{enumerate}
        \item An organization (a school, a bank)
        \item A particular professional (a doctor, a CPA)
        \item A friend, partner, or family member
        \item An employer or potential employer
        \item A landlord or potential landlord
        \item A government or governmental institution
        \item Other (please specify)
    \end{enumerate}
    \item How much did you trust the recipient (at the time you sent the information)? [Not well at all, moderately well, very well]
    \item Which of these scenarios most closely matches your experience with sending information like this using [METHOD]?
    \begin{enumerate}
        \item I suggested using [METHOD]
        \item The recipient suggested using [METHOD]
        \item The recipient only accepted information using [METHOD]
        \item I discussed with recipient and we jointly decided to use [METHOD]
    \end{enumerate}
    \item Please rate your agreement with the following statements. [Strongly disagree, disagree, neither agree nor disagree, agree, strongly agree, not sure]
    \begin{enumerate}
        \item I am tech-savvy.
        \item The recipient is tech-savvy. 
    \end{enumerate}
    \item How satisfied are you that using [METHOD] kept your data safe? [Very dissatisfied, dissatisfied, satisfied, very satisfied, unsure]
    \item What does it mean to you for your data to be safe? [free text]
    \item How satisfied are you with the convenience of using [METHOD] in the scenario you described? [Very dissatisfied, dissatisfied, satisfied, very satisfied, unsure]
    \item Please rate the likelihood of each of the following situations with regards to the scenario you described. [Very unlikely, unlikely, neither likely nor unlikely, likely, very likely, unsure]
    \begin{enumerate}
        \item The recipient will unintentionally reveal my data (e.g. a data breach, being hacked)
        \item The recipient will intentionally reveal my data (e.g. selling marketing information to a third party)
        \item My data will be intercepted in transit (someone other than the intended recipient will receive it)
    \end{enumerate}
    \item Please rate your agreement with the following statements: [Strongly disagree, disagree, neither agree nor disagree, agree, strongly agree, unsure]
    \begin{enumerate}
        \item The recipient has good procedures to keep my information safe.
        \item I can do something on my end to keep my information safe.
        \item This method is inherently secure
        \item This method ensures the information would be received by the intended recipient
    \end{enumerate}
    \item If it were your choice, would you use [METHOD] for sending sensitive information again? [Definitely would use again, would use again, would not use again, definitely would not use again, unsure]
    \item Please explain why you would or would not [METHOD] again. [free text]
    \item Please rate the risk of each of the following items in the scenario you have provided.  [No risk, low risk, high risk]
    \begin{enumerate}
        \item Identity theft.
        \item Monetary or financial harm.
        \item Reputational damage.
        \item Being targeted for harassment.
        \item Physical harm (including COVID-related concerns).
        \item The recipient may not receive my information.
    \end{enumerate}
    \item Please describe any other risks that could occur from the scenario you provided. [free text]
    \item Please explain your reasoning for rating [random selection of risk responses] as [no risk/low risk/high risk based on their response] [free text]
    \item Please rate the following statements regarding the scenario you have provided. [Strongly disagree, disagree, neither disagree nor agree, agree, strongly agree]
    \begin{enumerate}
        \item The data is at risk on my end of the transaction.
        \item The data is at risk while it is in transit.
        \item The data is at risk when it arrives at its destination.
    \end{enumerate}
    \item Please select the option that best describes your gender.
    \begin{enumerate}
        \item Male
        \item Female
        \item Nonbinary
        \item Another gender (please specify)
        \item Prefer not to say
    \end{enumerate}
    \item What is your annual income?
    \begin{enumerate}
        \item < \$50,000
        \item \$50,000 - \$100,000
        \item > \$100,000
        \item Prefer not to say
    \end{enumerate}
    \item What is your age? Please type "0" if you prefer not to say. [free text]
    
    \item Please choose the highest level of education you have completed.
    \begin{enumerate}
        \item Have not completed high school
        \item High school degree or equivalent
        \item Associate’s degree
        \item Bachelor’s degree
        \item Master’s degree
        \item Professional degree beyond a bachelor’s degree (e.g. MD, DDS)
        \item Doctoral degree
        \item Prefer not to say
    \end{enumerate}
    \item Do you have a computer science background? This means working in or holding a degree in computer science or information technology.
    \begin{enumerate}
        \item Yes
        \item No
        \item Not sure
        \item Prefer not to say
    \end{enumerate}
    \item Have you ever held a government security clearance?
    \begin{enumerate}
        \item Yes
        \item No
        \item Not sure
        \item Prefer not to say
    \end{enumerate}
    \item Have you ever handled any of the following information in a professional context? Please mark all that apply.
    \begin{enumerate}
        \item HIPAA
        \item FERPA
        \item Social security numbers
        \item Credit card data
    \end{enumerate}
    \item Is there any feedback on our survey or additional information you'd like to provide to help us understand your responses or improve the survey? [free text]
\end{enumerate}
}

\end{document}